\documentclass[preprint2]{aastex}
\usepackage{breqn}
\usepackage{bmpsize}

\slugcomment{submitted to ApJ}

\shorttitle{Structures in the UDF}
\shortauthors{Mei et al.}

\begin{document}


\title{Star-forming blue ETGs in two newly discovered galaxy overdensities in the HUDF at z=1.84 and 1.9: unveiling the progenitors of passive ETGs in cluster cores}


\author{Simona Mei\altaffilmark{1,2,3}, 
Claudia Scarlata\altaffilmark{4},
Laura Pentericci\altaffilmark{5},
Jeffrey A. Newman\altaffilmark{6},
Benjamin J. Weiner\altaffilmark{7},
Matthew L. N.  Ashby\altaffilmark{7},
Marco Castellano\altaffilmark{5},
Chistopher J. Conselice\altaffilmark{8}, 
Steven L. Finkelstein\altaffilmark{10} ,
Audrey Galametz \altaffilmark{5,14},
Norman A. Grogin\altaffilmark{9},
Anton M. Koekemoer\altaffilmark{9},
Marc Huertas--Company\altaffilmark{1,2},
Caterina Lani\altaffilmark{8,13}, 
Ray A.  Lucas\altaffilmark{9},
Casey Papovich\altaffilmark{11},
Marc Rafelski\altaffilmark{3,12} ,
Harry I. Teplitz\altaffilmark{3}
}


\altaffiltext{1}{GEPI, Observatoire de Paris, PSL Research University,  CNRS, University of Paris 
Diderot, 61, Avenue de l'Observatoire 75014, Paris France}
\altaffiltext{2}{University of Paris Denis Diderot, University of Paris Sorbonne Cit\'e (PSC), 75205 Paris Cedex
  13, France}
\altaffiltext{3}{Infrared Processing and Analysis Center, California Institute of Technology, Pasadena, CA 91125, USA}
\altaffiltext{4}{Minnesota Institute for Astrophysics, School of Physics and Astronomy, University of Minnesota, Minneapolis, MN 55455, USA}
\altaffiltext{5}{INAF, Osservatorio Astronomico di Roma, Via Frascati 33, 00040 Monteporzio, Italy}
\altaffiltext{6}{University of Pittsburgh, 3941 OÕHara St., Pittsburgh, PA 15260, USA}
\altaffiltext{7}{Steward Observatory, University of Arizona, 933 N. Cherry St., Tucson, AZ  85721}
\altaffiltext{8}{School of Physics and Astronomy, University of Nottingham, Nottingham, NG7 2RD, UK}
\altaffiltext{9}{Space Telescope Science Institute, Baltimore, MD, USA}
\altaffiltext{10}{The University of Texas at Austin, 2515 Speedway, Stop C1400, Austin, Texas 78712, USA}
\altaffiltext{11}{George P. and Cynthia Woods Mitchell Institute for Fundamental Physics and Astronomy, Texas
A\&M University, 4242 TAMU, College Station, TX 78743, USA}
\altaffiltext{12}{NASA Postdoctoral Program Fellow, Goddard Space Flight Center, Code 665, Greenbelt, MD 20771, USA}
\altaffiltext{13}{School of Physics and Astronomy, Tel Aviv University, Tel Aviv 69978, Israel}
\altaffiltext{14}{Max-Planck-Institut fur Extraterrestrische Physik (MPE), Postfach 1312, 85741 Garching, Germany}


\begin{abstract}
We present the discovery of two galaxy overdensities in the Hubble Space Telescope UDF:  a proto--cluster, HUDFJ0332.4-2746.6 at $z = 1.84 \pm 0.01$, and a group,  HUDFJ0332.5-2747.3 at $z =1.90 \pm 0.01$.  Assuming viralization, the velocity dispersion of
HUDFJ0332.4-2746.6 implies a mass of $M_{200}= (2.2  \pm 1.8) \times 10^{14}
M_{\sun}$, consistent with the lack of extended X--ray emission.
Neither overdensity shows evidence of a red sequence. About $50\%$ of their members show interactions and/or disturbed morphologies, which are signatures of merger remnants or disk instability. Most of their ETGs have blue colors and show recent star--formation. These observations reveal for the first time large fractions of spectroscopically confirmed star--forming blue ETGs in proto--clusters at $z\approx 2$.  These star--forming ETGs are most likely among the progenitors of the quiescent population in clusters at more recent epochs. Their mass--size relation is consistent with that of passive ETGs in clusters at $z\sim0.7-1.5$. If these galaxies are the progenitors of cluster ETGs at these lower redshifts, their size would evolve according to a similar mass--size relation. It is noteworthy that quiescent ETGs in clusters at $z=1.8-2$ also do not show any significant size evolution over this redshift range, contrary to field ETGs. The ETG fraction is  $\lesssim 50\%$, compared to the typical quiescent ETG fraction of $\approx 80\%$ in cluster cores at $z< 1$.  The fraction, masses, and colors of the newly discovered ETGs imply that other cluster ETGs will be formed/accreted at a later time.\end{abstract}


\keywords{Galaxy Clusters : general}



\section{INTRODUCTION}

Galaxy clusters are the largest structures observed in the
Universe. Their distribution and (baryonic and dark) matter content
constrain the cosmological model, and the study of their galaxy
properties reveals the influence of dense environments on galaxy
evolution.

Galaxies in clusters typically show a predominant early--type population 
and a red sequence  (old stellar population) up to redshift  z$\approx
1.5-2$ (e.g., Kodama et al. 2007; Mei et al. 2009; Andreon \&
Huertas--Company 2011;  Papovich et al. 2010; Snyder et al. 2012; Stanford et al. 2012;
Zeimann et al. 2012; Gobat et al. 2013; Muzzin et al. 2013; Mantz et al. 2014). Most of the clusters observed in the local Universe have assembled
their current early--type galaxy population at those redshifts (e.g., Cohn \&
White 2005; Li et
al. 2007; Chiang et al. 2013).
The redshift range around $z\approx 1.5-2$, however, has remained largely unexplored until
recently. The reason is that surveys based on cluster X--ray emission or the Sunyaev Zel'dovich effect (SZ) 
lack depth and/or area to reach detections of typical clusters at these redshifts, and 
ground--based optical spectroscopy would require excessive exposure times to confirm them spectroscopically when detected in the infrared/far--infrared bandpasses.

In the past five years, cluster samples at $z>1.5$ have been significantly enlarged by the
advent of deep and large enough surveys in the infrared and
mid-infrared, such as GOODS-MUSIC (Castellano et al. 2007), the IRAC Distant Cluster Survey (IDCS;
Eisenhardt et al. 2008; Stanford et al. 2012; Zeimann et al. 2012),
the Spitzer Deep, Wide-Field Survey (SDWFS; Ashby et al. 2009), the
Spitzer SPT Deep Field (SSDF; Ashby et al. 2013a), the Spitzer Adaptation of the Red-sequence Cluster Survey (SpARCS; Muzzin et al. 2013), Spitzer Wide-Area Infrared Extragalactic (SWIRE; Papovich et al. 2010), and the
Clusters Around Radio-Loud AGN program (CARLA; Galametz et al. 2012; Wylezalek et al. 2013). Other cluster candidates have been identified around low luminosity radio sources (Chiaberge et al. 2010). Spectroscopic
capability to confirm redshifts has been enhanced by optical and infrared grism spectroscopy with the Wide Field Camera 3
(WFC3)  on the Hubble Space Telescope (HST), and infrared
ground--based multi--object spectroscopy with the VLT/KMOS (Sharples
et al. 2006), the Keck MOSFIRE (McLean et al. 2010; 2012) and the
SUBARU MOIRCS (Ichikawa et al. 2006) instruments.

Until now, most clusters detected at $z>1.5$ have been identified as 
overdensities of red galaxies (e.g., Gladders \& Yee 2000), then confirmed by the spectroscopic follow--up of at least five members within 2~Mpc  (e.g., Castellano et
al. 2007, 2011; Kurk et al. 2009; Papovich et al. 2010; Tanaka et al. 2010; Stanford et al. 2012; Zeimann et al. 2012;
Muzzin et al. 2013), and/or by their X--ray emission (Andreon et al. 2009,
2011; Gobat et al. 2011; Santos et al. 2011). Four clusters have been
spectroscopically confirmed 
at z$\sim 1.8-2$: JKCS~041
(Andreon et al. 2009; Newman et al. 2013), IDCS J1426+3508 (Stanford
et al. 2012), IDCS J1433.2+3306 (Zeimann et al. 2012), and CL J1449+085 (Gobat et al. 2011; 2013). For all of the clusters with $z>1.8$,
spectroscopic redshifts have been obtained with grism spectroscopy
from HST/WFC3,  after ground--based optical spectroscopy failed to obtain enough signal.
These
systems show large fractions ($\approx 50\%$) of star--forming galaxies, indicating
that most of the quenching of star formation observed at lower
redshift had not yet occurred (Tran et al. 2010; Fassbender et al. 2011;
Hayashi et al. 2011;
Tadaki et al. 2012;
Zeimann et al. 2012; Brodwin et al. 2013). Recent observations at these redshifts also suggest that the
specific star formation of galaxies in dense regions becomes higher than
that in the field, although not all results are consistent with the supposed
reversal of the star-formation density relation (Elbaz et
al. 2007; Cooper et al. 2008; Gr{\"u}tzbauch et al. 2011; Hatch et al. 2011; Popesso et al. 2012; Andreon 2013; Gobat et al. 2013; Koyama et al. 2013;
Strazzullo et al. 2013; Santos et al. 2014; Scoville et al. 2013; Tanaka et al. 2013; Ziparo
et al. 2013).

 Current X--ray and SZ observations probe cluster
virialization through the detection of the hot gas in the gravitational potential well, down
to cluster masses of $\approx 10^{14} M_\odot$ and up to redshift of
$z \approx 1$. At higher redshifts, only the extreme end of the cluster mass function can be detected by current instruments.
A few objects at $1.5<z<2$ correspond to significant X--ray 
detections and were identified as already virialized
(Andreon et al. 2009; Gobat et al. 2011; Santos et
al. 2011; Stanford et al. 2012; Mantz et al. 2014). Two of them also show a significant SZ signal (Brodwin et al. 2012; Mantz et al. 2014). Their cluster masses cover the range of $M_{200}\approx (0.5 - 4)  \times 10^{14}
M_{\sun}$. The other detections (e.g., less massive objects) can only currently be identified as
significant (passive or active) galaxy overdensities, without confirmation of 
virialization by the detection of hot gas. Depending on the presence, 
or not, of the red sequence and their richness, these objects have been
identified as clusters or proto--clusters (e.g., Pentericci et
al. 2000; Miley et al. 2004, 2006; Venemans 2007; Kuiper et al. 2010; Hatch et al. 2011). In this paper, we will use the term proto--cluster to mean a cluster in formation, in agreement with this literature. In our definition, a cluster in formation, or proto--cluster, is either (1) a cluster that has not yet formed a red sequence, and, as a consequence, is detected as an overdensity of star--forming galaxies, or (2) a cluster that has not yet assembled and whose galaxies are distributed in groups that eventually will collapse to form a cluster (e.g. Chiang et al. 2013). Depending on the object richness/mass a galaxy overdensity is defined as group or cluster. Numerical simulations show that 90$\%$ of dark matter halos with masses  of $M_{200} \ge 10^{14} M_\odot$  are a very regular virialized population up to a redshift of $z\sim1.5$ (Evrard et al. 2008), and many works define as galaxy overdensities that have at least this mass as clusters. However, some other works define groups up to $ M \le few \ 10^{14} M_\odot $ (e.g. Yang et al. 2007), and the definition of galaxy overdensities as a group or a cluster varies in the literature. 

In this work, we will use the definition of clusters as overdensities with a mass of $M \ge 5 \times 10^{13} M_\odot$, because in previous studies of clusters at $z>1.5$ objects in this mass range have been defined as clusters in formation, or proto--clusters (e.g. Papovich et al. 2010). In fact, halos of this mass range at $z\sim1.5$ will be most probably accreted in clusters with masses of $M > 10^{14} M_\odot$ at $z<0.5$ (e.g. Chiang et al. 2013; Cautun et al. 2014).

Concerning the build--up of their early--type population, various studies have focused on the evolution of galaxies in clusters/dense environments from $z\approx 2$ to the present, and compared it to the field (Rettura et al. 2010; Cooper et al. 2012; Mei et al. 2012; Papovich et al. 2012; Raichoor et al. 2012; Bassett et al. 2013; Huertas--Company et al. 2013ab; Lani et al. 2013; Newman et al. 2013; Poggianti et al. 2013; Shankar et al. 2013; Strazzullo et al. 2013; Vulcani et al. 2013; Delaye et al. 2014; Shankar et al. 2014). These results indicate that the median/average passive ETG sizes in clusters are larger (within $\sim  2\sigma$),
and the analysis of the population with larger sizes suggests a different morphological type (E, S0) fractions and/or recently quenched faint galaxies.

In this paper, we present the discovery of two galaxy overdensities at
redshift of $z=1.84$ and $z=1.9$ in
the HST Ultra--Deep Field (HUDF; Beckwith et al. 2006) with
observations from  the Cosmic Assembly Near-infrared Deep
Extragalactic Legacy Survey (CANDELS; PI: S. Faber, H. Ferguson; Koekemoer et al. 2011; Grogin et
al. 2011), and the 3D HST survey (PI: P. van Dokkum; van Dokkum et
al. 2013; Brammer et al. 2012).  In Sec.~2, we present the observations. In Sec.~3 we describe our spectroscopic sample selection. In Sec.~4 we present the newly discovered overdensities and estimate one
 structure's mass. In Sec.~5, we study the stellar population and
 structural properties of their galaxies.  In Sec.~6 we conclude
and in Sec.~7 we summarize our results.

We adopt a $\Lambda CDM$ cosmology, with $\Omega_m  =0.3$,
 $\Omega_{\Lambda} =0.7$, and $h=0.72$.
All magnitudes are given in the AB system (Oke \& Gunn
1983; Sirianni et al. 2005). Stellar masses are estimated with a Chabrier initial mass function (Chabrier 2003).

\section{OBSERVATIONS}





The  Hubble Ultra-Deep Field (HUDF; Beckwith et al. 2006) is a 200$\arcsec \times $~200$\arcsec$ area  with the deepest HST observations in multiple wavelengths. HUDF has been observed by several
programs since the first HST Advanced
Camera for Surveys (ACS) data release in 2004,
including deep WFC3 images as part of the HUDF09 program (PI: G. Illingworth; Bouwens et
al. 2011), CANDELS, 3D-HST and HUDF12 (PI: R. Ellis; Ellis et al. 2013; Koekemoer et al 2013). 
CANDELS is a 902-orbit Multi-Cycle Treasury survey with the
HST, completed in Cycle~20.
The main instrument used by the survey is WFC3, with 3, 4 and 6
orbit exposures in imaging with the WFC3/F105W ($Y_{105}$), F125W ($J_{125}$), and F160W ($H_{160}$) filter, respectively, and grism
spectroscopy in the infrared (WFC3/IR) channel. Parallel observations were undertaken with the
ACS. 
A combination of all the HUDF observations with ACS and WFC3 has been
recently released by the eXtreme Deep Field (XDF)
program (Illingworth et al. 2013). We will use the combined XDF images
for the galaxy structural properties analysis, in particular imaging
with ACS/WFC (Wide Field Camera) F775W, F814W and F850LP  
($i_{775}$, $I_{814}$, $z_{850}$, respectively), for a total exposure time of
377.8~ks, 50.8~ks and 421.6~ks, respectively, and WFC3/IR $J_{125}$ and
$H_{160}$  for a total exposure time of 112.5~ks and 236.1~ks,
respectively.  The ACS WFC
resolution (pixel size) is 0.05\arcsec/pixel, and its  field of view is 202\arcsec x 202\arcsec. 
WFC3/IR has a 136\arcsec x 123\arcsec~field of view, with a spatial
resolution of 0.13\arcsec/pixel. 
The images have been drizzled and registered to obtain ACS and WFC3
mosaic images with the same resolution of 0.06\arcsec. The image
reduction is described in detail in Illingworth et al. (2013). We have verified that our results do not change when using the HUDF12 release (Koekemoer et al. 2013).

For the spectroscopy, the HUDF has been observed by two HST Treasury
programs with spectroscopic observations: the  CANDELS
and the 3D-HST  program. The 3D-HST program, completed in Cycle~19, obtained  deep spectroscopy of the HUDF with
the WFC3/IR G141 grism.  The grism spectroscopy from these two programs was recently
released as combined reduced spectra that include 8 orbits of 3D-HST
and 9 orbits of CANDELS supernova follow-up observations, for a total
of 17 orbits of observations (Brammer et al. 2012). 
The WFC3/IR G141 grism has an efficiency larger than
$30\%$ in the wavelength range $1.1<\lambda <1.65 \mu m$, a spatial resolution of 0.13\arcsec/pixel and a
dispersion of 46.5\AA/pixel. Typical uncertainties are 5{\AA} for
the zero point and 0.04{\AA} for the dispersion (Kuntschner et
al. 2010). The spectra
were extracted by independent software developed by the 3D-HST
collaboration, as described in Brammer et al. (2012), and redshifts
have been estimated using both grism spectroscopy and broadband photometry for a
combined spectro-photometric estimate. For the entire spectroscopic
catalog, spectral features used to estimate redshifts include rest--frame
H$\alpha$, [O~II]$\lambda$3727,  [O~III]$\lambda$5007 emission lines, and
the Balmer 4000{\AA} break. The 3D-HST spectroscopy covers an area of $\sim 140\arcsec \times 140\arcsec$ in the HUDF.

Near ultra--violet images (NUV) of the HUDF were obtained in a Hubble Space
Telescope treasury program (hereafter UVUDF; Teplitz et al. 2013) using the WFC3/UVIS detector. This project obtained deep images of the
HUDF in the F225W, F275W, and F336W filters.
Data were obtained in two observing
modes (as described in Teplitz et al. 2013), with $\sim 15$ orbits of integration per filter in each mode.
 For the current analysis,
we use the half of the data that were obtained with the post-flash
(the UVIS
capability to add internal background light), to mitigate
the effects of degradation of the charge transfer efficiency of the
detectors (Mackenty \& Smith 2012). The data were reduced using a
combination of standard and custom calibration scripts (see Rafelski et al. 2014, in prep.), including the use
of newly released software  to correct for charge transfer
inefficiency.  The individual reduced exposures were 
then registered and combined following the methods developed
for CANDELS (Koekemoer et al. 2011). The 5$\sigma$\ rms
sensitivities in an aperture with 0\farcs2 radius are 27.9, 27.9, and
28.3~mag in $F225W$, $F275W$, and $F336W$, respectively. Photometry in the UV was
measured in isophotal areas determined from the B-band detection image
obtained with SExtractor in dual image mode (Bertin \& Arnouts
1996).

\section{SPECTROSCOPIC SAMPLE SELECTION}

Using the CANDELS and 3D-HST spectroscopic
redshifts, we identified an initial galaxy
overdensity in the HUDF at redshift $z \approx 1.85$.  We explain below how we identified and quantified this detection. 

To assess the quality of the spectra, we used both visual
inspection, the published spectro--photometric analysis
from the 3D-HST collaboration (Brammer et al. 2012), and the
CANDELS Guo et al. (2013) photometric redshift catalog. For 3D-HST spectroscopy, we applied the
shift $z_{spec}=0.005 \times (1+z_{3DHST})$, as suggested in the
documentation of the 3D-HST data release (van Dokkum et al. 2013, 3D-HST data release documentation). The accuracy
of the Guo photometric redshifts is estimated to be
$\delta_{z_{pz}} = 0.030 \times (1 + z)$ for $H_{160}<24$~mag and
$\delta_{z_{pz}} = 0.039 \times (1+z)$
for $H_{160}>24$~mag, with a global outlier fraction of $\approx 4 \%$. This gives typical photometric redshift errors
of $\sigma_{pz} = 0.09 - 0.1$ at z=1.8-1.9 up to $H_{160} \approx
26$~mag. The Guo et al. catalog covers the HUDF area ($\sim$4.6~$arcmin^2$), and
  extends to the CANDELS/GOODS-S field, in the deep
  ($\sim 55~arcmin^2$) and wide ($\sim 30~arcmin^2$)  CANDELS surveys. The Guo et
  al. (2013) photometric catalog has a $5~\sigma$ magnitude depth of 27.4, 28.2, and
  29.7 AB, for an aperture of 0.17'', in the CANDELS wide,
deep, and HUDF fields, respectively.

We (BW first and then SM verified and agreed) flagged each spectrum as (1) certain, (2) good independently of
photometric redshift estimates,  (3) good using photometric redshift
estimates, (4) probable, and (5) not usable. For this work, we only use certain and good spectroscopic grism redshifts (flag 1 to 3).
We have been particularly
attentive to the possible contamination from misidentification of
H$\alpha$ emission as O~III, from the foreground cluster at $z$=1.096 (Salimbeni
et al. 2009), e.g., we have not considered two galaxies 
because they show a single line emission and their photometric redshifts would indicate a most probable
redshift at $z \sim 1$. The lines detected with significant signal--to--noise (S/N)
ratio are specified in Table~1, and are mainly
[O~III]$\lambda$5007 or the O~III doublet, and [H$\beta$]$\lambda$4861. 

We found 24
galaxies, which all lie within a radius of $R=2${\arcmin} (that corresponds to a comoving
radius of 1~Mpc at $z\approx1.8-1.9$) of the main overdensity center (see below), and have good
quality spectra from which we measure redshifts in the range
$1.8<z<1.95$.  In this range, the grism redshift median statistical
uncertainty is
$\approx 0.001$ (Brammer et al. 2012; Colbert et al 2013). To the
statistical uncertainty, we add a systematic of $0.003 \times
(1+{z_{spec}})$. We estimate the systematics from the median scatter when comparing spectroscopic redshifts measured by
the CANDELS collaboration (BW) with spectro-photometric redshifts published  by the 3D-HST collaboration (Brammer et al. 2012), in the redshift range $z=0.5-2.5$. 
This systematic is larger than, but consistent with, the uncertainties
obtained from the simulations by Colbert et al. (2013) and exactly the same as found by Gobat et al. (2013).

We searched the entire GOODS CDF-S master catalog
(\footnote{http://www.eso.org/sci/activities/garching/projects/goods/MasterSpectroscopy.html})
for spectroscopy from ground--based  follow-up
of the area. When grism
redshifts are probable or not usable, and 
ground--based multiple line redshift measurements are available with
average S/N per pixel $>3$ (Kurk et al. 2013), we use
VLT/FORS2 redshift measurements instead of the grism spectroscopy. Eight
galaxies with redshifts in the range of
$1.8<z<1.95$ have good quality GMASS VLT/FORS2 spectroscopy and respect
our S/N criteria, and four have better quality than the grism
spectra. We added the missing 4 to the 24 galaxies above, to obtain 28 galaxies with
good quality spectra in the range of 
$1.8<z<1.95$. The 3D-HST spectra of the cluster members were published by the 3D-HST collaboration (Brammer et
al. 2012). The GMASS spectra were published in Kurk et
al. (2013).  

The selected spectroscopic members extend to magnitudes as faint as $H_{160}
\approx 25.7$~mag; however, the grism spectroscopy sample shows a
marked decrease in number at magnitudes fainter than $H_{160} \approx
24.5$~mag. At this magnitude, $\approx 95\%$ of the CANDELS  galaxies  in
the HUDF area (Guo et al. 2013) have a grism redshift estimation, and $\approx 50\%$ have
good quality flags from our classification above.

The UVUDF NUV images permitted us to confirm that the selected galaxies are at $z\sim1.8-1.9$,  since we expect them to be UVIS/$F225W$
dropouts (see Teplitz et al. 2013)  and to be detected in the $F336W$ filter. Each candidate has been inspected visually and independently by two of us (CS and
SM). Two of the candidates,
UDF--1898 and UDF--1909 are $F336W$ dropouts,  and could either be galaxies at higher redshift or too faint to be detected. We will not consider these two objects in the rest of our
analysis, leaving 26 selected galaxies with redshift of $1.8<z_{spec}<1.95$.
In the appendix, we show the WFC3 $F225W$, $F275W$,
$F336W$, and ACS $F435W$, $I_{814}$ and WFC3 $J_{125}$ images for each candidate.

We describe in Table~1 all of the selected galaxies, and identify them by their 3D-HST UDF ID (Brammer et al. 2012) or GMASS ID (Kurk et al. 2013).


\begin{figure*}
\epsscale{1.5}
\center{ \includegraphics[width=0.8\textwidth,natwidth=610,natheight=642]{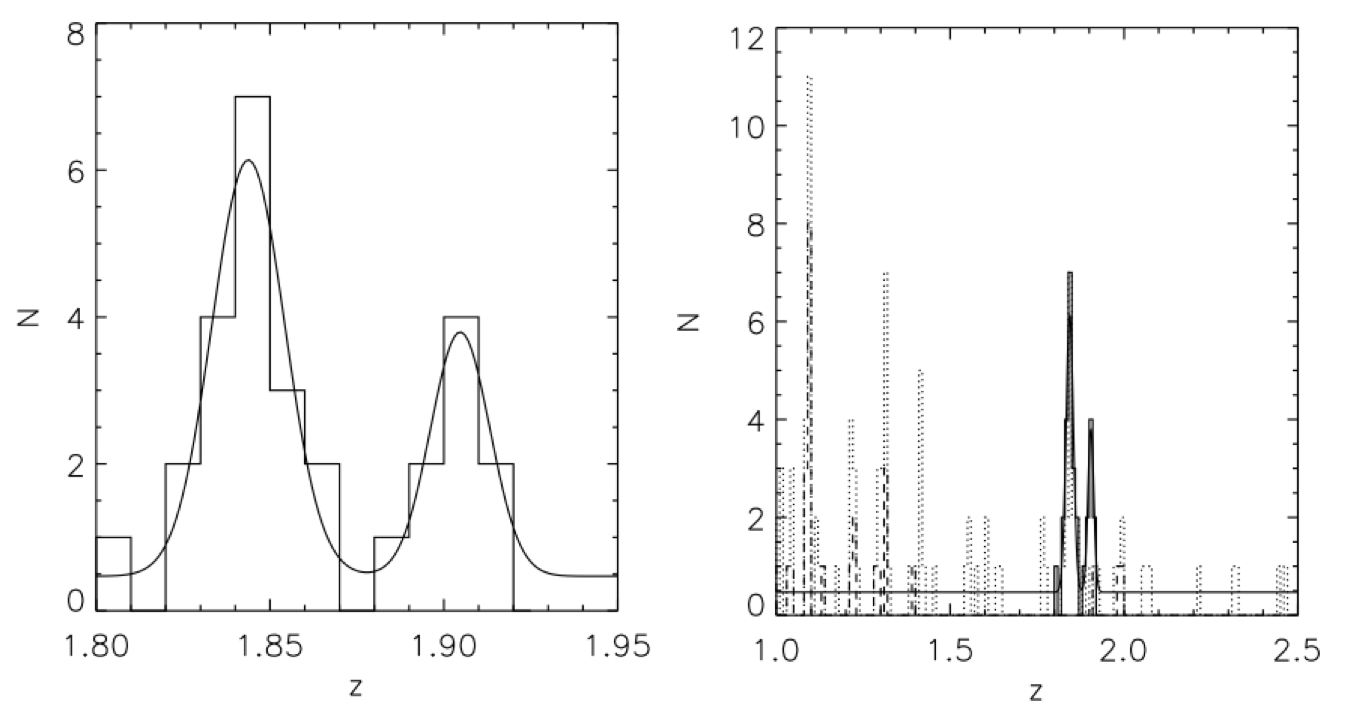}}
\caption{On the left, the redshift distribution of HUDFJ0332.4-2746.6 and
  HUDFJ0332.5-2747.3 spectroscopic 
   members. The continuous line is the double Gaussian fit described in the text. On the right, we also show the spectroscopic redshift distribution in the HUDF from the 3D-HST (dashed line) and the GOODS and GMASS catalogs (dotted line).\label{histo}}
\end{figure*}

\section{NEWLY DISCOVERED OVERDENSITIES IN THE HUDF}

\subsection{Structure definition using spectroscopy}

As shown in Fig.~\ref{histo}, the redshifts of the 26 selected galaxies appear to follow a double Gaussian distribution.  This
is confirmed by a
classical Kolmogorov-Smirnov test, skewness and kurtosis test and the
two more robust asymmetry index (A.I.) and tail index (T.I.) described
in Bird \& Beers (1993; as in e.g., Castellano et al. 2011). The tests were applied on each Gaussian separately  (e.g. we only considered redshifts within 3$\sigma$ from each mean for each test). For the structure at $z=1.84$, we obtained a skewness of $0.05 \pm 0.10$, a kurtosis of $-0.49 \pm 0.17$, a T.I. of $1.4 \pm 0.1$ and an A.I. of $-0.2 +/- 0.5$. For the structure at $z=1.9$, we obtained  a skewness of  $-0.9 \pm 0.4$,  a kurtosis of  $-0.5 \pm 0.9$, an A.I. of $-0.4 \pm 0.5$ and the T.I. could not be calculated with so few points. We estimated uncertainties with a Monte Carlo simulation. All parameters are consistent with a random population extracted from a Gaussian distribution (Bird \& Beers 1993).

To select the structure members, given the small number of galaxies,
we fit a double Gaussian plus a background to the
observed distributions. We took into account Poissonian uncertainties in the
histogram and adopted a redshift bin of 0.01. The Gaussian fits give $\overline z = 1.84 \pm 0.01$ and $\overline z = 1.905 \pm 0.005$, for
the first and second structure, respectively. We obtain $\overline z =
1.84 \pm 0.01$ and $\overline z = 1.90 \pm 0.01$, respectively, from a biweight mean redshift and
standard deviation. 

Selecting galaxies within  $3 \times \sigma_{\overline z}$ of the two means, we obtained 18 spectroscopic
members for the first Gaussian and 7 spectroscopic members for the
second. One galaxy is not selected as part of the
structures because its redshift is too low. 

 To better estimate the significance of the two redshift overdensities, since we do not have a large area, we will measure statistics in redshift bins in the HUDF area covered by the grism spectroscopy, that corresponds to a co--moving size of $\sim 1$~Mpc. We
used the complete sample of certain and good (as defined above) spectroscopic  grism
redshifts from CANDELS and 3D-HST in the range $1.2<z<2$, and calculated both projected densities
using Nth-nearest neighbor distances and galaxy overdensities, following Papovich et al. (2010) (see also Gobat et al. 2013).  In this redshift range, WFC3 redshifts are mainly obtained from H$\alpha$ and O~III emission lines combined with photometric redshifts as explained above. While it is true that to have a precise estimate of the overdensities, we would need to use spectroscopic samples at the same redshift over a large area,  such a sample is currently not available. However, even if galaxies at different redshift have redshift estimates based on different emission lines, e.g. H$\alpha$ and O~III, and the flux limit increases with redshift, these two effects would point to a lower limit for our O~III emission line overdensities, since (1)  they would be at the higher redshift end of the range in redshift that we considered and (2) they are defined by their O~III emission and the O~III emission has similar or lower  strength than 
H$\alpha$  (e.g. Colbert et al. 2013). We did not consider GMASS redshift measurements in this estimation, because our sample is dominated by emission line galaxies. In both calculations, we have considered all galaxies brighter than $H_{160}=27$~mag.

For the first overdensity estimate, we measure projected densities
using Nth-nearest neighbor distances defined as $\Sigma_N=\frac{N}{\pi
  D^2_N}$ (e.g., Dressler et al. 1980). $N$ is the number of neighbors, $D_n$ is defined as the distance in Mpc to the Nth nearest
neighbor. We have calculated $\Sigma_N $ within redshift bins of
amplitude 0.06 (e.g., within a distance in redshift space $3 \times \sigma_{\overline
  z}$ from the biweight analysis) from $z= 1.2$ to $z=2$. The significance of our detections is estimated by taking the ratio: $S/N = \frac{\Sigma_N - \Sigma^{bck}_N}{\sigma_{\Sigma^{bck}_N} }$.Our background density estimates were stable in the range
$N=3-7$, with $\Sigma^{bck}_N=0.5 \pm 1$. The structure at $z=1.84$ is an
overdensity at $\approx 20 \sigma$ above the background density
(stable for $N=4-7$, it is  $\approx 14 \sigma$ at $N=3$). 
The structure at $z=1.9$ has a density at $6-8 \sigma$ above the background
density for $N=3$ and 4, respectively. Given the smaller number of galaxies, this measurement is less stable at different $N$. Our results do not change if we enlarge the redshift range, and do not consider in the analysis the known cluster at $z=1.096$ (see above).

For the second overdensity estimate, we use the definition of galaxy contrast $\delta_{c}=\frac{N_{gal}-N_{bkg}}{N_{bkg}}$. $N_{gal}$ is the number of galaxies in a given redshift bin, and $N_{bkg}$ is the average number of background galaxies in the entire redshift range of $1.3<z<2$. The significance of our detections is estimated by taking the ratio $S/N = \frac{N_{gal}-N_{bkg}}{\sigma_{bkg}}$. We obtain $ N_{bkg}= 0.6 \pm 0.9$ galaxies 
per redshift bin of 0.06. Even if we do not count $> 3 \sigma$ peaks in the redshift distribution, this might be an upper limit to the average background, since we already know that there are significant overdensities in this field  (Salimbeni et al. 2009).   For the first and second structure, we obtain a $\sim18\sigma$ and a $\sim6\sigma$  galaxy overdensity, respectively. These results are consistent with those from projected densities
using Nth-nearest neighbor distances.

This analysis confirms the detection of the structure at
$z=1.84$ as a significant galaxy overdensity. We call this structure
HUDFJ0332.4-2746.6, and adopt as the center the position of its
brightest galaxy (UDF~2095, with $H_{160} = 22.008 \pm 0.002$~mag) in its spatially denser
region of comoving size $R=500$~kpc (1{\arcmin} at this redshift), at
[RA, DEC]=[53.155647,-27.779298].  Only one member (GMASS~220) is farther than
$R=1\arcmin$ from this center. The measured overdensity is similar to that measured for red galaxy overdensities that were confirmed as galaxy clusters by their X--ray emission (Papovich et al. 2010; Gobat et al. 2011). Also, numerical simulations of our standard cosmological model (Hanh et al. 2007a, 2007b; Cautun et al. 2014) predict that galaxy overdensities of $\sim 30$ at $z\sim2$ have a probability of $\sim40\%$ of being a node of the cosmic web (e.g. a cluster or proto-cluster) and a probability of $\sim10\%$ of being a filament. While predictions from Cautun et al. (2014) and Hanh et al. (2007a, 2007b) do not take into account projection effects, long filaments (we would need a filament with a comoving lenght of 50~Mpc seen in projection in the area of HUDFJ0332.4-2746.6 to reproduce our measured overdensity)  at $z\sim2$ account for only a very small percentage ($<10\%$ from Cautun et al. 2014) of the total finalement length distribution. The number of filaments decreases with the filament length, and dense and long filaments are also rare with a probability of $\sim2 \times 10^{-4}$ of being found in the HUDF area (fig. 53 from Cautun et al. 2014). Therefore, HUDFJ0332.4-2746.6 has a higher probability of being a cluster of galaxies than a filament.

\begin{figure*}
\epsscale{.80}
\plotone{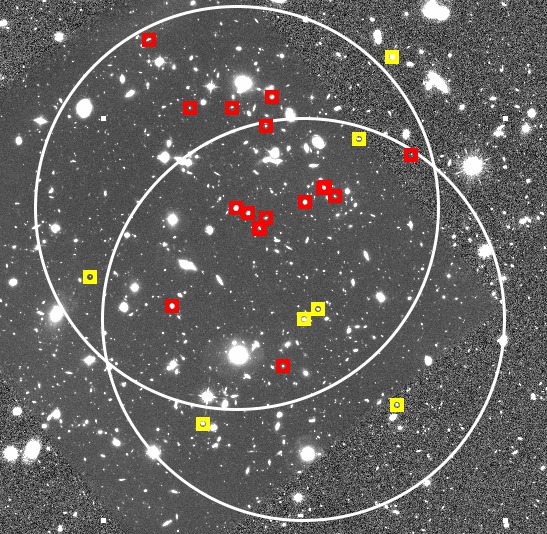}
\caption{Our structures' galaxies. Over the $H_{160}$ image of the HUDF, the red and yellow boxes show spectroscopic 
  members for HUDFJ0332.4-2746.6 and HUDFJ0332.5-2747.3, respectively. Both UDF-2090 and UDF--2103, and UDF--2433 and UDF--2491,
   are pairs and cannot be distinguished in the figure. The large white
   circles, centered on each structure, have a radius of $R=1\arcmin$, that corresponds to a comoving radius of $\approx 0.5$~Mpc.
   Most of HUDFJ0332.4-2746.6 are within $1\arcmin$  from the
   structure center, the only exception is
   GMASS220. HUDFJ0332.5-2747.3  is more sparse. North is on
  the top, east on the left.  \label{region}}
\end{figure*}

The second overdensity is  detected at  $ 4-7\sigma$, and we will call it HUDFJ0332.5-2747.3.  We adopt as the center the position of  the
brightest galaxy ($H_{160} = 22.463 \pm 0.002$~mag) in its spatially denser
region, at
[RA, DEC]=[53.149298 ,-27.788534].   Numerical simulations of our standard cosmological model (Hanh et al. 2007a, 2007b; Cautun et al. 2014) predict that galaxy overdensities of $\sim 10$ at $z\sim2$ have a probability of $<10\%$ of being a node of the cosmic web (e.g. a cluster or proto-cluster) and a probability of $\sim50\%$ of being a filament. We identify this structure as a galaxy group because it is less populated and less compact, and its detection threshold is closer to that of a galaxy group (e.g., Tanaka et al. 2013), and it also has a high probability of being a filament according to numerical simulations.

The positions of the structures' members are shown in Fig.~\ref{region}. The comoving distance between the two  structures is $\sim 100 $~Mpc. At $z\sim 2$, the standard cosmological model predicts that the comoving volume of progenitors of clusters with present masses $M>10^{14} M_{\sun}$ can reach $\sim 25$~Mpc$^3$ for the most massive clusters (Chiang et al. 2013; Shattow et al. 2013), thus the two structures are not predicted to necessarily merge to form a present--day cluster.

\begin{figure*}
\epsscale{1.8}
\plotone{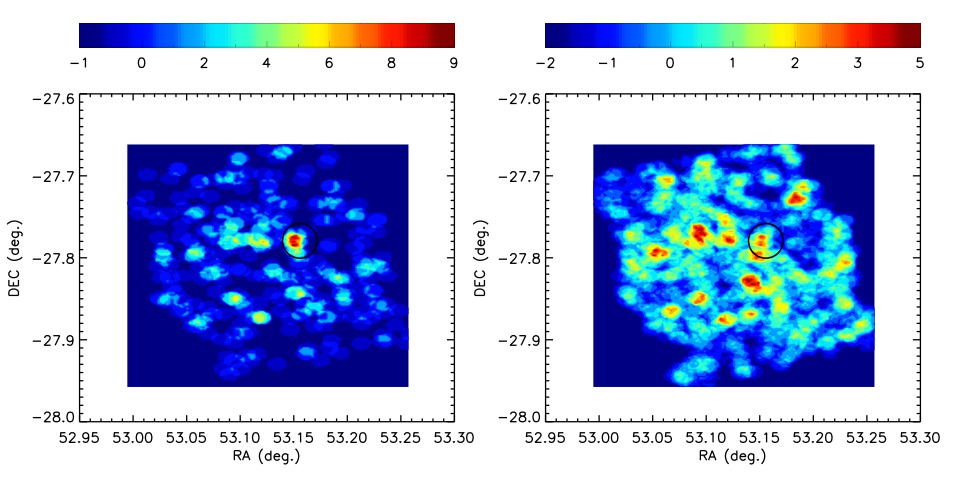}
\caption{Photometric redshift overdensities at the same redshift as our spectroscopically detected overdensities (see text for the exact photometric redshift range). On the left and the right, we show overdensities obtained at a depth of $H_{160}=24.5$~mag and  $H_{160}=26$~mag, respectively. The black circles are centered on the HUDFJ0332.4-2746.6 center, and have comoving radius of $\sim$0.5~Mpc. In both cases, our structures are found as one single overdensity, and are part of a larger overdense structure that extends over a region of $\sim$12~$arcmin^2$~($\sim$6~Mpc$^2$, in comoving distance). \label{fig_pzod}}
\end{figure*}

\subsection{Photometric redshift overdensities}

 While  the CANDELS and 3D-HST spectroscopic covers only the HUDF area, the Guo et al. (2013) photometric redshift catalog covers the entire GOODS-S field for a total area of $\sim$170~$arcmin^2$. This means that when using photometric redshifts, we can extend our overdensity search over a comoving projected area of $\sim 80$~Mpc$^2$ at $z=1.8-19$. We will use these measurements to investigate if our overdensities are isolated or are part of a larger--scale overdensity distribution at the same redshift.

\begin{figure*}
\epsscale{1.3}
\plotone{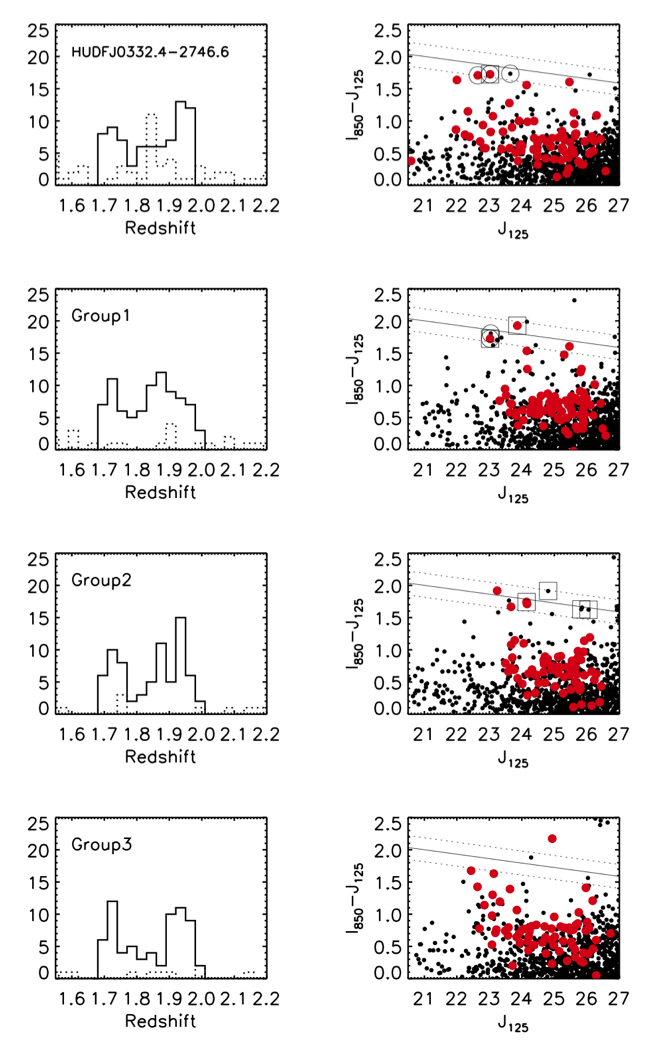}
\caption{On the right, we show the histogram of the photometric redshift members (continuous line, $1.69<z_{phot} < 1.99$) within 1.5\arcmin ~from the center of each photometric redshift overdensity in Table~\ref{table_pzod}. The dotted histogram shows the distribution of galaxies with known spectroscopic redshifts. On the left, we show the color--magnitude relation for each overdensity. The $(I_{850} - J_{125})$ color is close
  to the $(U-B)$ rest--frame, and $J_{125}$ to the B--band rest--frame at $z\sim1.84$. The black points are all galaxies within 1.5\arcmin~from the overdensity center. The larger red points are the galaxies with $1.69<z_{phot} < 1.99$. The squares and circles around symbols indicate an AGN detection and a known spectroscopic redshift, respectively. The continous line shows the color--magnitude relation at $z\sim1$ from Mei et al. (2009) passively evolved at z=1.84, and the dashed lines show a region within 3 times the observed scatter. Some of the overdensities show a red sequence, even if none of the red sequence galaxies with known spectroscopic redshifts has spectroscopic redshifts within 3$\sigma$  of HUDFJ0332.4-2746.6 and HUDFJ0332.5-2747.3 spectroscopic redshift.  \label{cmr-ods}}
\end{figure*}

\begin{figure*}
\epsscale{1.3}
\plotone{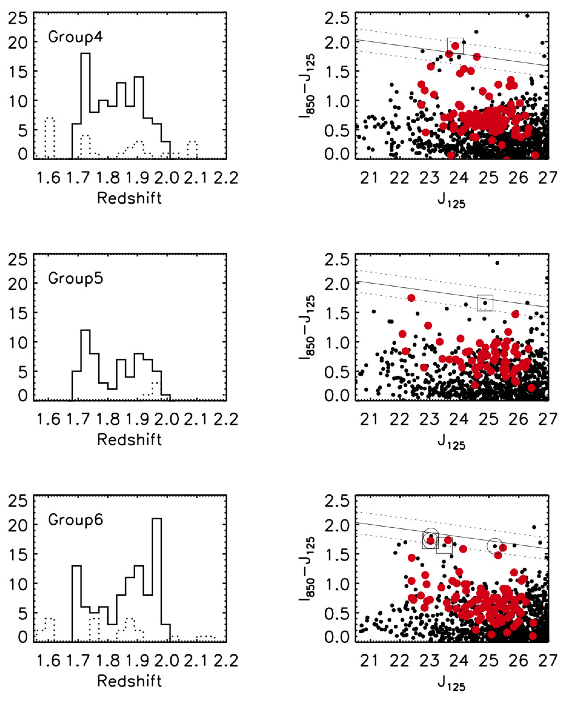}
\caption{Same as Fig.~\ref{cmr-ods}, but for the remaining groups from Table~\ref{table_pzod}. \label{cmr-ods2}}
\end{figure*}

We have selected all galaxies with magnitudes brighter than $H_{160} = $~24.5, and 26~mag, which correspond to median uncertainties on the single photometric redshifts of $\lesssim$~0.1 and 0.15, respectively. Given the larger uncertainties associated with photometric redshifts, we measured overdensities in photometric redshift ranges at $\pm 1 \sigma$ from the center of our larger overdensity, HUDFJ0332.4-2746.6. Given the uncertainties on photometric redshift we cannot separate HUDFJ0332.4-2746.6 from HUDFJ0332.5-2747.3 in this analysis. The redshift ranges that we considered for sample depths of $H_{160} = $~24.5 and 26~mag, are of $1.74<z_{phot}<1.94$~mag and $1.69<z_{phot}<1.99$~mag, respectively. In this last photometric redshift range, we estimated the purity and completeness of our photometric redshift catalog using the spectroscopic sample. We find it to be $\sim 70\%$ complete and $\sim 60\%$ pure.

We identified overdensities in regions of a projected comoving radius of 0.5~Mpc, as $\delta_{c}=\frac{N_{gal}-N_{bkg}}{N_{bkg}}$, and define $S/N = \frac{N_{gal}-N_{bkg}}{\sigma_{bkg}}$. Fig.~\ref{fig_pzod} shows the S/N of the overdensities that we detected, and we list the most significant overdensities in Table~\ref{table_pzod}, selecting all overdensities with the three highest signal--to--noise ratios in each of two cases, and to always include HUDFJ0332.4-2746.6. At both depths, HUDFJ0332.4-2746.6 and two overdensities, that we called Group~1 and Group~2 are detected within the three highest signal--to--noise ratios. As we expect from the median photometric redshift uncertainties, when we increase the depth in magnitude we also increase the background and the background noise, and our detections are less significant. We considered as a single detection all overdensities closer than a comoving distance of 0.5~Mpc. Group~1 is at a comoving distance of $\sim$1~Mpc from HUDFJ0332.4-2746.6, and Group~4. UDFJ0332.4-2746.6 and HUDFJ0332.5-2747.3, with Group~1, Group~4 and Group~6 form a structure that extends extends to $\sim 10$\arcmin $\times 10$\arcmin~($\sim 5 \times 5$~comoving Mpc). All Groups together cover a region of $\sim 12$\arcmin $\times 12$\arcmin~($\sim 6 \times 6$~comoving Mpc). 
 
 \begin{table*}
\begin{center}
\caption{Photometric redshift overdensities \label{table_pzod}}
\vspace{0.25cm}
\resizebox{!}{2.5cm}{
\begin{tabular}{llccccccccccccccc}
\tableline \tableline\\
Name & $RA$ (deg.)&$DEC$ (deg.)&$H_{160}^{lim}$&$N_{gal}$&$N_{spec}$, $\overline z_{spec}$&$S/N$&$R$(arcmin.)\\
 \tableline \tableline\\
HUDFJ0332.4-2746.6& 53.15565  & -27.77930 &24.5&8&&7& -\\
Group~1& 53.11842   &-27.78338 &24.5&6&5,$1.89\pm0.01$&5&1.9\\
Group~2&53.11615 &  -27.87192  &24.5&6&--&5&5.9\\
Group~3&    53.19252  & -27.82862    &24.5&5&3,$1.88\pm0.03$&4&3.5\\
\tableline \\
HUDFJ0332.4-2746.6& 53.15565  & -27.77930 &26&13&&2&-\\
Group~1& 53.11842   &-27.78338 &26&13&5,$1.89\pm0.01$&2&1.9\\
Group~2& 53.11615  & -27.87192  &26&14&--&2&5.9\\
Group~4& 53.09392  & -27.76772 &26&18&6,$1.88\pm0.02$&3&3.3\\
Group~5&    53.18884&   -27.72558    &26&15&4,$1.95\pm0.01$&2&3.7\\
Group~6& 53.14208  & -27.81992   &26&14&8,$1.87\pm0.02$&2&2.5\\
 \tableline \tableline\\
 \end{tabular}}
\end{center}
\small{
This Table shows properties of the photometric redshift overdensity in the GOODS-S field, with photometric redshifts consistent with the HUDFJ0332.4-2746.6 and HUDFJ0332.5-2747.3 spectroscopic redshifts. $H_{160}^{lim}$ is the magnitude limit we have used in our overdensity search, $RA$ and $DEC$ are the position of the overdensity center, $N_{gal}$ the number of galaxies selected within 1\arcmin~for each overdensity, $N_{spec}$ the number of galaxies with good quality spectroscopic redshifts in the range $1.8<z_{spec}<2$, and their average value, S/N is the overdensity significance, as defined in the text, $R$ the distance from  the HUDFJ0332.4-2746.6 center. Group~1 and Group~4, and Group~1 and Group~6 have one galaxy in common, respectively.
}
\end{table*}

In Fig.~\ref{fig_pzod}, we show the overdensities. In Fig.~\ref{cmr-ods} and Fig.~\ref{cmr-ods2}, we show the photometric and spectroscopic redshift histograms for each overdensity. In Table~\ref{table_pzod}, we give the number of galaxies with $1.8<z_{spec}<2$ within 1.5\arcmin~from each overdensity center, and their average spectroscopic redshift. From the available spectroscopy, Group~1, Group~3, Group~4 and Group~6 have three to eight galaxies that show an average spectroscopic redshift for each structure close to HUDFJ0332.4-2746.6 and HUDFJ0332.5-2747.3 (see Table~\ref{table_pzod}). All of these structures together show 19 galaxies within 3$\sigma$ (observational scatter) from the red sequence measured in Mei et al. (2009) at $z\sim1$, when it is passively evolved at $z=1.84$. Group~5 has a higher average spectroscopic redshift from four galaxies, at $\overline z_{spec} = 1.95\pm0.01$. Given the few galaxies that are spectroscopically confirmed, we cannot analyze the groups in detail.

If confirmed as significant spectroscopic overdensities, some of the galaxies in these groups might be part of the same large--scale structure as HUDFJ0332.4-2746.6 or HUDFJ0332.5-2747.3, but without extensive spectroscopic follow--up, we cannot draw a firm conclusion. As thoroughly discussed in the literature (e.g. Shattow et al. 2013 and references therein), projection effects can strongly affect fixed aperture measurements of overdensities, especially when using high uncertainties in photometric redshifts. Our detections have to be confirmed by spectroscopic follow--up for a better quantification of their significance.

\subsection{HUDFJ0332.4-2746.6 Mass Estimate}

\subsubsection{X-ray observations}

We checked the 3 Msec XMM and 4 Msec Chandra X--ray observations for
both point sources associated with the galaxies in
the two overdensities and for possible extended emission from the ICM.
\\
The HUDFJ0332.4-2746.6 member UDF--2095 coincides with source \# 512
in the 4 Msec catalog (Xue et al. 2011). It has a soft band flux of
	$1.8\times 10^{-17} erg/sec/cm^2$ and a hardness ratio of 0.27.
The catalog classifies the emission as a {\it galaxy}
so it is most probably associated with star formation  rather
than with AGN activity.
There are other X--ray sources within the cluster region, of which one (\# 505 from Xue et al. 2011) is
extended but associated with a lower redshift galaxy (z=0.99).
There is no indication of diffuse extended emission coinciding with
either overdensity position.
\\
From the lack of extended X--ray emission, we can place an upper
limit on the X--ray luminosity of 1-6$\times10^{43} erg/s$,  depending on the temperature assumed (in the range $T$=1-3 KeV, respectively). If we
use the cluster mass--luminosity derived by Rykoff et al. (2008),  this
corresponds to an upper limit in total mass of $M_{200} < 1 \times 10^{14}
M_\odot$, and $M_{200} < 3 \times 10^{14}
M_\odot$, for an upper limit in the X--ray luminosity of 1 and 6~$\times10^{43}$ erg/s, respectively. 
This means that we cannot exclude that the most massive structure is a cluster with mass $M_{200} \sim 10^{14} M_\odot$.

\subsubsection{Mass estimates}

From numerical simulations, we know that 90$\%$ of the halos with masses of $M_{200} \ge 10^{14} M_\odot$ up to $z\sim1.5$ have virialized (Evrard et al. 2008). Since we cannot exclude this hypothesis from our X--ray measurements (see the previous section), and an overdensity of galaxies at $\sim 20 \sigma$ can correspond to a halo of mass $M_{200} \sim 10^{14} M_\odot$ (e.g. Gobat et al. 2013), we decide to make this assumption. This is also supported by the fact that the velocity distribution of HUDFJ0332.4-2746.6 is consistent with a Gaussian, e.g. it has already separated from the Hubble flow  (e.g.
Nakamura 2000; Merrall \& Henriksen 2003). 

With this assumption, HUDFJ0332.4-2746.6 mass can be  
estimated from its velocity dispersion. 
The line--of--sight (LOS) cluster velocity dispersion can be highly anisotropic, and 
small samples lead to large systematic uncertainties (White et
al. 2010).  For  HUDFJ0332.4-2746.6, we expect uncertainties in the velocity dispersion from anisotropies of $\approx 10\%$. We do not  do the same for HUDFJ0332.5-2747.3, in fact, the uncertainty on the mass estimate for HUDFJ0332.5-2747.3 is too large because of the smaller number of galaxies, and we do not attempt to measure it.

We measure the HUDFJ0332.4-2746.6 intrinsic velocity dispersion
from its 18 members, following Danese et al. (1980).
We add in quadrature the statistical and systematic uncertainties in redshift. From the overdensity intrinsic velocity dispersion, we obtain an estimate of the mass using Eq.~(1) from the $\Lambda CDM$ simulations in Munari
et al. (2013): 

\begin{equation}
M_{200}=\left( \frac{\sigma_{1D}}{A_{1D}} \right)^{1/\alpha} \frac{10^{15} M_{\sun}}{h(z)}
\end{equation}
with the parameters $A_{1D} = 1090 \pm 50$, and $\alpha = 0.3333$ (see
also Evrard et al. 2008). $\sigma_{1D}$ is the cluster  LOS velocity dispersion
$\sigma_{disp}$, and $h(z)=H(z)/(100 km/s)$, where $H(z)$ is the Hubble constant.
Assuming virialization, this equation gives the relation between the total mass of a cluster
in a radius $R_{200}$\footnote{ $R_{200}$ is the radius at which the 
cluster mean density is 200 times the critical density.} and its
velocity dispersion, and is obtained by using a Navarro, Frenk \&
White (1996) dark matter mass profile with different concentration
parameters and different constant velocity anisotropies. The
uncertainty in the coefficient $A_{1D}$ takes into account the
uncertainties in these simulation assumptions. 
From the cluster velocity dispersion we also derive $R_{200}$  (Carlberg et
al. 1997).

\begin{figure*}
\epsscale{1.8}
\plotone{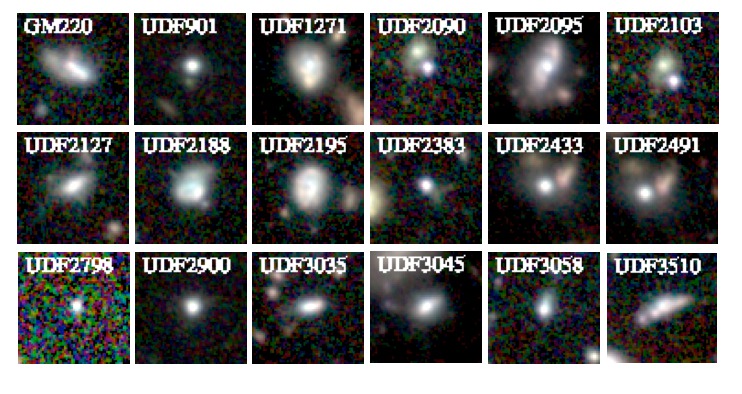}
\caption{HUDFJ0332.4-2746.6 : combined color image of the
  spectroscopic members, from the ACS
  $B_{435}$, WFC3 $i_{775}$ and $H_{160}$ images. 
Most of the galaxies
  present spiral morphologies, as expected from most
  star--forming galaxies. Eight are classified as ETGs (see Table~\ref{tab1}). Most show asymmetries, faint substructures and tails, which are
  signatures of merger remnants. Four galaxies form two confirmed
  pairs, others have close small companions that are not confirmed to
  be at the same redshift. \label{members1}}
\end{figure*}

Following the classic Danese et al. (1980) computation of the intrinsic velocity dispersion, and its uncertainty, we obtain $\sigma_{disp}=780_{-100}^{+180} $~km/s. This corresponds to a mass of $M_{200}=2_{-1}^{+2}  \times 10^{14}
M_{\sun}$, and  $R_{200} = 1.0_{-0.1}^{+0.3}$~Mpc. 

To take into account possible systematics due to the sample selection, we estimate the uncertainty on the cluster velocity
dispersion, its mass, and its virial radius by bootstrapping 1,000
times on the 18 cluster members.  Specifically, we recalculated the three
quantities, $\sigma_{disp}, M_{200}$, and $R_{200}$, substituting all the initial sample with a sample of the same size extracted randomly from the initial sample. We obtain an intrinsic velocity dispersion of $\sigma_{disp}=(730\pm260) $~km/s, $M_{200} = (2.2 \pm 1.8) \times 10^{14} M_{\sun}$, and  $R_{200} = (0.9 \pm 0.3)$~Mpc. This suggests that using the classic computation from Danese et al. (1980) does not take into account all uncertainties in the sample selection, and we will use these last estimates as more robust. If we underestimated systematics on redshift measurements, our mass estimate becomes an upper limit. 

When using other values of $A_{1D}$, obtained using two different models of the baryonic
 physics in  Munari et al. (2013), our results do not significantly change. A systematic of $\approx 10\%$ in velocity
dispersion from the LOS anisotropies would lead to a
systematic of $\approx 10-15\%$ in mass, in this range of velocity
dispersion and mass, and does not change our results.

From our conservative result $M_{200} = (2.2 \pm 1.8) \times 10^{14} M_{\sun}$, we can derive a simple quantification for the probability of our hypothesis of virialization. We are consistent with a mass of $M_{200} \ge 1 \times 10^{14} M_{\sun}$ at $\sim 75\%$ of probability. If we assume that the structure  just started to separate from
the Hubble flow (Steidel et al. 1998), we would obtain a mass of the same order of magnitude ($\approx 10^{14}
M_{\sun}$), but with larger systematics due
to the difficulty in estimating the three--dimensional volume that the overdensity covers with the
available low resolution spectroscopy. With the available data, this is the best that we can do.

\begin{figure*}
\epsscale{1.8}
\plotone{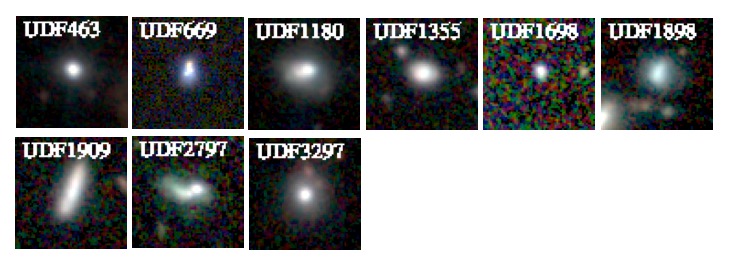}
\caption{HUDFJ0332.5-2747.3 : Combined color image of the
  spectroscopic members, from the ACS
  $B_{435}$, WFC3 $i_{775}$ and $H_{160}$ images. 
Most of the galaxies
  present spiral morphologies, Five have early--type morphologies. \label{members2}}
\end{figure*}

\section{HUDFJ0332.4-2746.6 AND HUDFJ0332.5-2747.3 GALAXY POPULATION}
\subsection{Morphologies}

All of the structures' galaxies but two (see Table~1) show recent star formation. In Fig.~\ref{members1} and Fig.~\ref{members2}, we show their color images.  Most of the
morphologies are disturbed, and often show asymmetry and/or
asymmetric tails. Using the  WFC3
$J_{125}$ imaging, which corresponds to the B
rest--frame at z=1.84-1.9,  we visually classified galaxies into two categories: ETGs
and late--type galaxies (LTGs). We included compact galaxies in the ETG
class, and irregular galaxies in the LTG class. We have seven ETGs at
z=1.84, plus UDF--3058 that we consider a ETG with an asymmetric tail,
and five ETGs at z=1.9. 
UDF-1355 looks
like an ETG in the B--band rest--frame but shows asymmetric features in the UV. We consider as reliable only the two ETGs brighter than $H_{160} =
23.5$~mag (van der Wel et al. 2012; Kartaltepe et al. 2014).
LTGs always reveal structure and are therefore all reliably
classified (see also, e.g., Mortlock et al. 2013; Kartaltepe et al 2014). 

Among the candidates at $z=1.84$, UDF--3058 has an
ambiguous morphology, with a bulge--like appearance and an
asymmetric tail that appears in the B rest--frame. We classify it as
an ETG. One galaxy that we classify as an ETG, UDF~2900, shows a
double core in the UV.

For the 17 galaxies with $H_{160} <
24.5$~mag, our classification is consistent with the CANDELS
morphological classification from Kartaltepe et al. (2014) for all galaxies. This is
a higher than typical level of consistency among different
classifiers/classification methods when the morphological classification
includes only two broad classes, ETGs and LTGs  (e.g., Postman et al. 2005;
Huertas--Company et al. 2009, 2011).

Two spectroscopic pairs are close companions, but
we do not have enough spectral resolution to 
identify them as mergers. From Kartaltepe et al. (2014), two objects
are classified as mergers ($12^{+13}_{-8}$~\%), seven as interacting
($41 \pm 14$~\%), six as asymmetric ($35^{+15}_{-13}$~\%), four have tidal
  features  ($23^{+14}_{-10}$~\%). UDF~3297 has been marked as having a double
    nucleus. All of this accounts for nine objects  ($53^{+14}_{-15}$~\% of the sample), because some objects have multiple features.  It is interesting that half of
    these galaxies are interacting or disturbed, because they show signatures that are characteristic of merger remnants or disk instability (see also results from, e.g., Lotz et al. 2013; Mortlock et al. 2013).The galaxies in these structures are still being assembled and the observation of interactions and disturbed morphologies point to mergers and possibly disk instabilities as the primary mechanisms.
    
The fraction
of confirmed early--type galaxies is at
most $48 \pm10 \%$ of the entire sample, against the typical $\approx 80\%$ and close to the $\approx 50-80\%$
observed in galaxy clusters at $z<1.5$ and $1.5<z<2$, respectively, for galaxy masses of $M>10^{10-10.5} \times M_{\sun}$ and a total cluster mass of $M>10^{14} \times M_{\sun}$ (e.g., Postman et
al. 2005; Desai et al. 2007; Mei et al. 2009, 2012;  Tran et al. 2010; Fassbender et al. 2011;
Hayashi et al. 2011; Papovich et al. 2012;
Tadaki et al. 2012;
Zeimann et al. 2012; Brodwin et al. 2013).  Our results have to be taken as upper limits to the fractions of ETGs in our structures. In fact, we emphasize that we only consider as secure ETGs those galaxies with $H_{160} <23.5$~mag, e.g., two over the five ETGs. This means that when calculating the fraction of ETGs, we might be overestimating it, since the three fainter ETGs might not be ETGs.

\begin{figure*}
\center{\includegraphics[angle=0,scale=0.45]{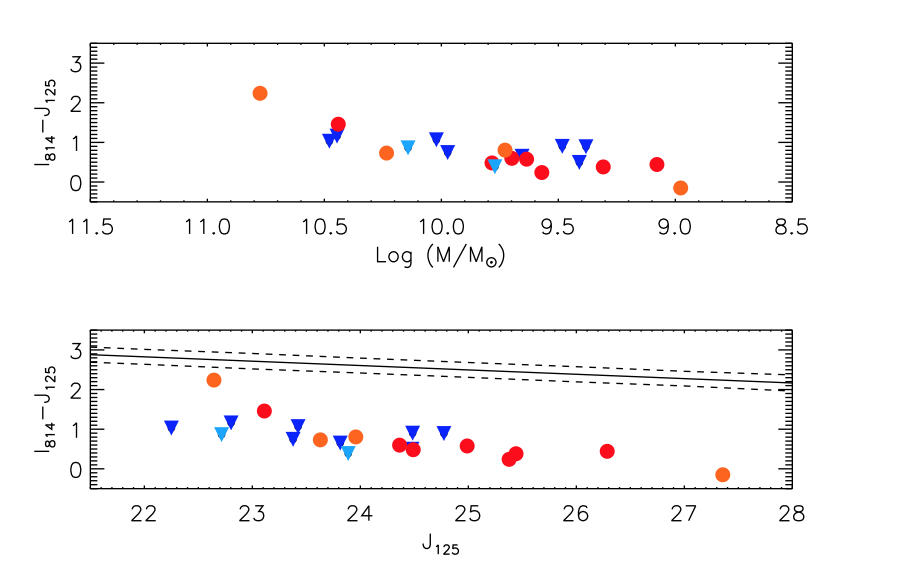}}
\caption{Color--magnitude and color-mass diagram for all
  HUDFJ0332.4-2746.6 and HUDFJ0332.5-2747.3 spectroscopically
  confirmed members with Guo et al. (2013) photometry. The selected galaxies have masses in the range $8.9 < log_{10}
  (\frac{M}{M_{\sun}}) < 11$. The $(I_{814} - J_{125})$ color is close
  to the $(U-B)$ rest--frame, and $J_{125}$ to the B--band rest--frame. Red/orange circles and blue/sky triangles are
   ETGs and LTGs in
  HUDFJ0332.4-2746.6/HUDFJ0332.5-2747.3, respectively. We show as a continuous line the color--magnitude relation at $z\sim1$ from Mei et al. (2009) passively evolved at z=1.84, and the dashed lines show a region within three times the observed scatter. A red sequence is not yet
  formed. \label{cmrmass}}
\end{figure*}

\subsection{Galaxy Masses and Colors}

We have used the Guo et al. (2013) photometric catalog to estimate
galaxy masses and colors. This catalog includes observations from
CANDELS HST/WFC3 $Y_{105}$, $J_{125}$ and $H_{160}$ data, combined with
existing public data from the HUDF09 programs. In addition to WFC3 bands, the catalog also includes data from UV (U-band
from both CTIO/MOSAIC and VLT/VIMOS), optical
(HST/ACS F435W, F606W, F775W, F814W, and F850LP), infrared (HST/WFC3
F098M, VLT/ISAAC Ks, VLT/HAWK-I Ks),  and Spitzer/IRAC 3.6, 4.5, 5.8,
8.0~$\mu m$ observations (from the GOODS and SEDS surveys: Fazio et al. 2004; Ashby et al. 2013b). We refer to Guo et al. (2013) for a detailed description of these observations.
The catalog is based on source detection in
 $H_{160}$, and all photometry was matched using the public
software TFIT (Laider et al. 2007). The photometry reaches
a 5$\sigma$ depth (within an aperture of radius 0.17{$\arcsec$}) of 
29.7~mag in the HUDF region, with a completeness of 50\% at 28.1~mag in
$H_{160}$.

We estimated galaxy masses from the Guo et al. broadband photometry, using the public software
Le~Phare (Arnouts et al. 2002; Ilbert et al. 2006), based on a 
$ \chi^2$ spectral energy distribution (SED)  fitting method.
For our Le~Phare input parameters, we
followed Ilbert et al. (2010) and used the Chabrier IMF, Bruzual \& Charlot (2003) templates, solar metallicity, an exponentially
decaying star formation with $\tau$ in the range 0.1--5~Gyr, and a
Calzetti et al. (2000) extinction law with E(B--V) in the range 0--0.5. 

The galaxies in our structures have magnitudes in the range of $22.3 \lesssim J_{125} \lesssim27.4$~mag and $22 \lesssim H_{160} \lesssim 26.3$~mag ($20.75 \lesssim H^{VEGA}_{160} \lesssim 25$~mag). Their range in luminosity is similar to magnitudes observed in the clusters detected at z=1.8-1.9 by Stanford et al. (2012) and Zeimann et al. (2012), even if these two clusters' most luminous galaxies are brighter of $\sim 0.5-1$~mag with respect to our most luminous galaxies. This difference of $\sim 0.5-1$mag is of the same order of magnitude as the difference in luminosity between the most luminous galaxies in different confirmed clusters at $0.8<z<1.3$ (Mei et al. 2009).

In Fig.~\ref{cmrmass} we show the color--magnitude and
color--mass relations. 
  All galaxies have masses in the range of 
$8.9 \lesssim log_{10} (\frac{M}{M_{\sun}}) \lesssim 10.8$, and all of their colors, but
one (UDF--463), which also does not show emission lines, are bluer than
quiescent galaxies at these redshifts. In fact, at z=1.84 and according to a Bruzual \&
Charlot (2003) simple, single starburst model with solar metallicity, we
would expect a red sequence at 
$(I_{814}-J_{125}) \approx 2.3$~mag, for a formation
redshift $z_f = 2.5$. This value corresponds to the mean luminosity-weighted formation redshift usually
derived for galaxies in clusters at $z\approx1-1.5$ (e.g., Mei et
al. 2006ab; Mei et al. 2009, 2012; Snyder et
al. 2012; Brodwin et al. 2013; and references therein). We show as a continuous line the color--magnitude relation at $z\sim1$ from Mei et al. (2009) passively evolved to z=1.84. The dashed lines show three times their total observed scatter ($ 3 \times \sigma_{obs} \sim 0.2$). Hereafter, we define as {\it red} galaxies those that are redder than the passively evolved red sequence minus 3$ \times \sigma_{obs}$.  It is clear that most of the ETGs in these structures still need to be quenched.

Unlike the known clusters at $z\approx 1.8-2$
(Stanford et al. 2012; Zeimann et al. 2012; Newman et al. 2013; Gobat et al. 2011, 2013) that
show overdensities of $\sim 10-15$
red galaxies, the two overdensities do not show an already formed red
sequence. Only one of the structure galaxies has a color red enough to be considered as a red sequence galaxy at these redshifts.

Potential red sequence galaxies could have been missed in our spectroscopical analysis because we only selected
star--forming galaxies, or only those with good quality spectra from
the 3D-HST, CANDELS and GMASS catalogs. However, when using the entire
Guo et al. photometric catalog in the HUDF, there are only six
  other red galaxies (e.g., as defined above, with $(I_{814}-J_{125})
>1.7$~mag) within $1.5\arcmin$ from the proto--cluster and group
centers from the spectroscopic redshift catalog from GMASS, the
photometric and photometric redshift catalog (used without any
selection in redshift) from Guo et al. (2013), the CANDELS morphology
catalog from Kartaltepe et al. (2014), which we have examined one by
one using the Guo et al (2013) photometry. The three brightest of the six red galaxies are within $1\arcmin$ from the proto--cluster, have
 spectroscopic redshift measurements that are at lower or at higher redshift, and do not belong to the two overdensities. 

We robustly confirm that HUDFJ0332.4-2746.6 and HUDFJ0332.5-2747.3 do not have an already formed red sequence, within $1.5\arcmin$ of the proto--cluster and group centers. 

We also examined the colors of the photometric redshift overdensities around HUDFJ0332.4-2746.6 and HUDFJ0332.5-2747.3. In Fig.~\ref{cmr-ods} and Fig.~\ref{cmr-ods2}, Group~1, Group~2, Group~4 and Group~6 have promising bright red sequences. Using the spectroscopic redshift catalog available in the GOODS area (Wuyts et al. 2008, 2009, in preparation; Kurk et al., in preparation) the red sequence galaxies with known redshifts (a circle surrounds their symbols in Fig.~\ref{cmr-ods} and Fig.~\ref{cmr-ods2}) do not show spectroscopic redshifts in the the range of HUDFJ0332.4-2746.6 and HUDFJ0332.5-2747.3. In the figures, we also show AGN detections from Xue et al. (2011). Group~1 has two red galaxies with spectroscopic redshifts, with $z=2.35$ and $z=1.76$. Group~6 has three red galaxies with spectroscopic redshifts, the two also belonging to Group~1 and one at higher redshift.

\begin{figure*}
\epsscale{1.5}
\plotone{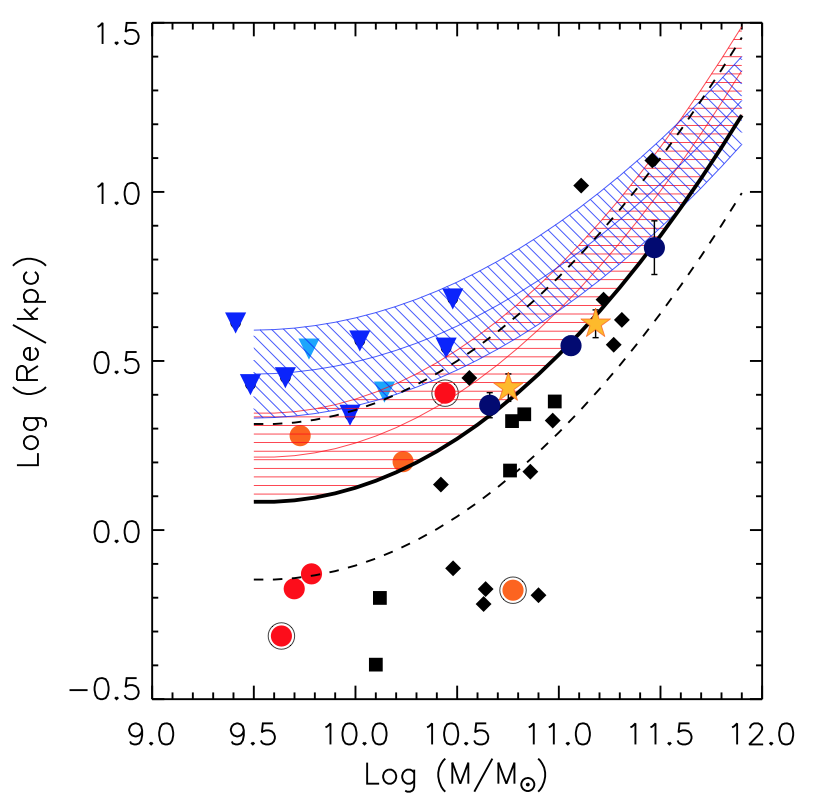}
\caption{Mass--size relation for galaxies with $H_{160} >
  24.5$~mag in HUDFJ0332.4-2746.6 and HUDFJ0332.5-2747.3. Red/orange circles and blue/sky triangles are
   ETGs and LTGs in
  HUDFJ0332.4-2746.6/HUDFJ0332.5-2747.3, respectively. The black circles around filled symbols show galaxies with $n>2$.
The two stars show the ETG mass-size relation observed in galaxy clusters at a 
redshift of $1.2<z<1.5$ from Delaye et al. (2014), and the dark blue circles show
results from Lani et al. (2013) ($1<z<2$), for their most dense regions. The filled squares and diamonds are the quiescent ETG masses and circularized effective radii from CL~J1449+085 at $z=1.99$ (from Strazzullo et al. 2013) and JKCS~041 at $z=1.8$ (from Newman et al. 2013), respectively. All ETGs from CL~J1449+085  and JKCS~041 have been selected as galaxies with $n>2$. As a reference, we show the SDSS local mass--size relation from Bernardi et al. (2012). The red/blue continuous line shows the mass--size relation for SDSS ETG/LTG, respectively. The shaded regions show the 1$\sigma$ observed scatter.  The continuous black line shows the local mass---size relation scaled to the average sizes from Delaye et al. at $10^{11} M_{\odot}$ (the dashed lines indicate the observed scatter). Our blue star--forming ETGs lie on the same mass--size relation as quiescent ETGs in dense environments at $1<z<2$. \label{massize}}
\end{figure*}

\begin{figure*}
\epsscale{1.5}
\plotone{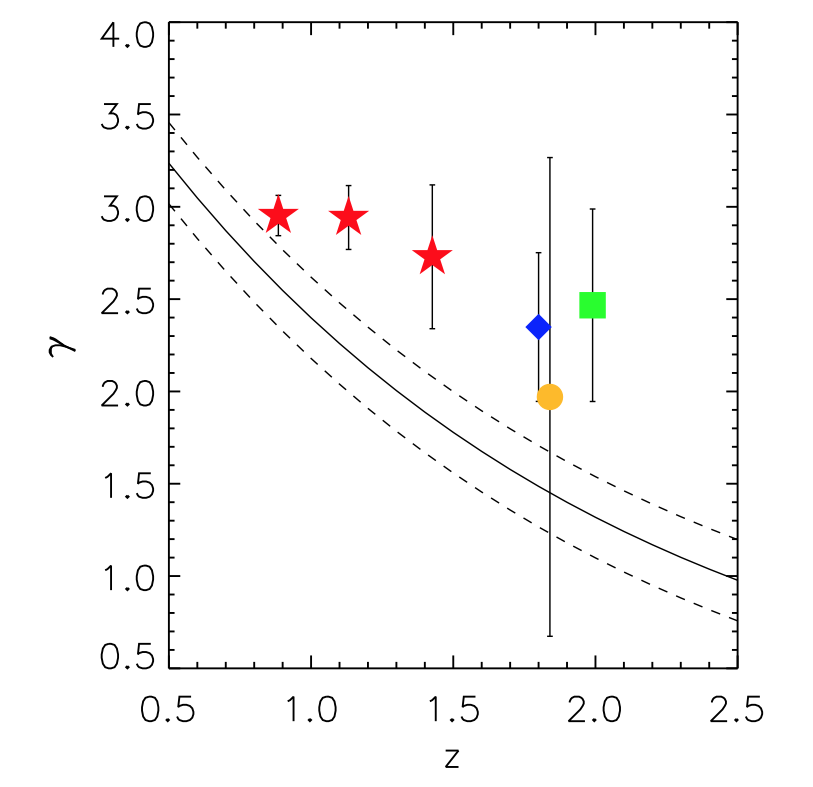}
\caption{Median mass-normalized B--band rest-frame size $\gamma$, as a function of redshift. The yellow circle is the median $\gamma$ for HUDFJ0332.4-2746.6 ETGs. The red stars are the median $\gamma$s for quiescent ETGs in clusters at $0.7<z<1.6$
from Delaye et al. (2014), The green square and the blue diamond are the median $\gamma$s for  quiescent ETGs from CL~J1449+085 at $z=1.99$ (Strazzullo et al. 2013) and JKCS~041 at $z=1.8$ (Newman et al. 2013), respectively. Our structures' star--forming blue ETGs are consistent with those of quiescent ETGs in dense environments at similar redshifts. Both star--forming and quiescent ETGs in dense environments do not show much evolution in the redshift range of $z=0.7-2$.  The continuos line shows the evolution of field galaxies and is from Newman et al. (2012).  \label{gevol}}
\end{figure*}

\subsection{Galaxy Structural Properties}

As demonstrated by van der Wel et al. (2012), the WFPC3/IR camera
resolution together with the depth of CANDELS observations,  permit us
to estimate galaxy structural parameters up to $H_{160}
\approx 23$~mag, and galaxy sizes  up to $H_{160}
\approx 24.5$~mag. Basset et al. (2013) have also shown that the same
applies up to $J_{125} = 24$~mag.

We estimated galaxy  structural parameters for all galaxies (ETGs
and LTGs) using GALFIT  (Peng et
al. 2002; 2010) on the WFC3/IR  $J_{125}$ image from XUDF, that
corresponds to the B--band rest--frame at $z\sim1.8-1.9$.
We have adopted a single Sersic profile, and did not constrain the values of the Sersic index $n$ for most of the
sample, as suggested in Peng et al. (2002). The  point spread function (PSF) model
was provided by van der Wel et al. (2012). The galaxies classified as late--type all have a Sersic index of $n<2$, the two bright ETGs have $n>2$ and the other ETGs have $1<n<2$. As the size estimate, we use the circularized effective radius $R_e$, defined as the average half-light radius along the major axis of the best--fitting galaxy model multiplied by the ratio between the minor and major axis $q=\sqrt{b/a}$.

Fig.~\ref{massize} shows the galaxy mass--size relation. 
The two stars show the ETG mass-size relation observed in galaxy clusters at
a redshift of $1.2<z<1.5$ from Delaye et al. (2014), and the dark blue circles show
results from Lani et al. (2013) ($1<z<2$), for their most dense regions (see
also Papovich et al. 2012; Basset et al. 2013). 

These previous works 
pointed out that ETGs in clusters have, on average, larger sizes than ETGs
in the field at the same redshift (within $\sim 2\sigma$), when the mass--size relation is
taken into account (see also Cooper et al. 2012; Raichoor et
al. 2012; Newman et al. 2013).
For clusters in the same redshift range that our structures, the filled square and diamonds are the ETG masses and circularized effective radii from CL~J1449+085 at $z\sim2$ (Strazzullo et al. 2013) and JKCS~041 at $z=1.8$ (Newman et al. 2013), respectively, with masses corrected to a Chabrier IMF. All sizes are measured in the B--band rest--frame.
 Our structures' ETGs all have masses of $M < 10^{11} M_{\odot}$, in the same range of the masses of passive ETGs in CL~J1449+085 (see the filled squares), and are about an order of magnitude lower than the most massive ETGs in the most massive cluster known at these redshifts, JKCS~041 (see the filled diamonds).

 As a reference, we show the SDSS mass--size relation for ETG and LTG galaxies from Bernardi et al. (2012). While the Bernardi et al. mass--size relation has been estimated for field galaxies, it also holds for galaxy cluster for the ETGs (Huertas--Company et al. 2013b), and we do not expect large variations for the LTGs (Fernandez--Lorenzo et al. 2013). Our structures' LTGs lie on the same mass--size relation as Bernardi et al. (2012) LTGs.  For the ETGs, assuming that the form of the mass--size relation from Bernardi et al. (2012) does not evolve with redshift, when extrapolating the Delaye et al. (2014) and Lani et al. (2013) mass--size relations at lower masses, our structures' ETGs follow the same mass-size relation at $1<z<2$.
 
 It is very interesting, because this is also true for the ETGs in JKCS~041 and CL~J1449+085. 
To better quantify this point, in Fig.~\ref{gevol}, we plot the mass-normalized B--band rest-frame size, $\gamma$, as a function of redshift, for passive ETGs in clusters at $0.7<z<1.6$ from Delaye et al. (2014), JKCS~041 at $z=1.8$ (from Newman et al. 2013) and CL~J1449+085 at $z\sim2$ (from Strazzullo et al. 2013), and the star--forming ETGs in HUDFJ0332.4-2746.6 from this work. Since our galaxies span a large range in mass ($10^{9.5} M_{\odot} \lesssim M \lesssim 10^{12} M_{\odot}$, we calculate $\gamma$ using the SDSS mass--size relation in Bernardi et al. (2012; Eq. 1), instead of the commonly used power law that holds for galaxy masses $M>10^{11} M_{\odot}$:
\begin{eqnarray}
log(\gamma) &=&  log(R_e)+c1 \times \left[log\left( \frac{ M}{10^{11} M_{\odot}}\right) \right] + \\
&&c2 \times \left\{ \left[log(M) \right]^2 -\left[log(10^{11} M_{\odot}\right)]^2 \right\} \nonumber
\end{eqnarray}
where $R_e$ and $M$ are the galaxy circularized effective radius and mass in units of $M_{\odot}$, respectively, and $10^{11} M_{\odot}$ is the typical mass used for the mass normalization. $c1$ and $c2$ are the coefficients for ETGs and LTGs from Bernardi et al. (2012).
The uncertainties have all been estimated by bootstrap with replacement, with 1000 iterations. Both the quiescent and the star--forming ETG median normalized sizes do not evolve significantly from $z\sim2$ to $z\sim0.7$ ($\sim 20\%$). 
When using the average $\gamma$ instead of the median, results are consistent. This redshift range corresponds to a time interval of $\sim 4$Gyrs, over which ETG sizes must have evolved on average according to the same mass---size relation as that of cluster ETGs at $z\sim1$.

 On the low--mass--end side ($M<10^{11}M_{\odot}$), our structures's ETGs must have had their star formation quenched, though, to be selected as passive ETGs in the $z\approx1$ samples.  

In Table~\ref{tab1}, we give the galaxy magnitudes, colors, masses, and structural parameters.

\begin{table*}
\begin{center}
\caption{Spectroscopic members for 
  HUDFJ0332.4-2746.6 and HUDFJ0332.5-2747.3\label{tab1}}
\vspace{0.25cm}
\resizebox{!}{4cm}{
\begin{tabular}{llccccccccccccccc}

\tableline \tableline\\
ID & RA (J2000)&DEC(J2000)&z &$H_{160}$
(mag)&$(I_{814}-J_{125})$(mag)& Morphology&Spectral Features (Q)&$Log_{10}(M/M_{\odot})$&$R_e \times \sqrt{b/a}(kpc)$&$N_{Sersic}$\\
\tableline \tableline \\
HUDFJ0332.4-2746.6 &&&&&\\
\tableline \\
             UDF-901 &        53.15126 &       -27.79241 &        1.841  $\pm$       0.001  &     23.970  $\pm$ 0.005    & 0.597  $\pm$       0.042&ETG&O~III,         H$\beta$ ,{\it H$\gamma$}  &  9.7&0.67 $\pm$ 0.01&1.85$\pm$ 0.04   \\
          UDF-1271/GMASS-675 &        53.16166 &       -27.78743
          &  1.836  $\pm$       0.008&  22.475  $\pm$ 0.002   & 1.158
          $\pm$       0.027 &LTG&{\it O~III, H$\beta$ }    (3.3)   &  10.5&3.45 $\pm$ 0.01&0.47$\pm$ 0.01   \\ 
      UDF-2090 &        53.15346 &       -27.78098 &        1.849  $\pm$       0.001&      24.489    $\pm$       0.008      &       0.576    $\pm$       0.071
&ETG    & O~III,    {\it H$\gamma$ }  &  9.6&0.49 $\pm$ 0.07&10$\pm$ 2          \\
          UDF-2095/GMASS-858 &        53.15565 &       -27.77930 &        1.839  $\pm$       0.003&      22.008  $\pm$       0.002          & 1.029  $\pm$       0.019  &   LTG&     O~III,    H$\beta$, H$\gamma$  (8.4)   &  10.5&4.82 $\pm$ 0.04&1.15$\pm$ 0.01      \\
UDF-2103 &       53.15351 &       -27.78091&       1.838    $\pm$ 0.004 &      & 0.576  $\pm$       0.071   &ETG&O~III    &  &&         \\         
        UDF-2127 &        53.15287 &       -27.78012 &       1.858  $\pm$       0.003&      24.292  $\pm$       0.008         &0.899  $\pm$       0.068 &    LTG &   O~III   &  9.5&2.69 $\pm$ 0.04&1.46$\pm$ 0.02            \\
         UDF-2188/ GMASS-875 &        53.15452 &       -27.77972 &       1.836  $\pm$       0.004  &      23.624  $\pm$       0.005         &0.649  $\pm$       0.035 &  LTG&      O~III      (5.1)   &  9.7&2.82 $\pm$ 0.02&0.43$\pm$ 0.01    \\
UDF-2195/GMASS-894 &        53.14921 &       -27.77880 &        1.850  $\pm$       0.002 &      23.189  $\pm$       0.004      & 1.064  $\pm$       0.038 &LTG&O~III, {\it O~IIIx,
  H$\beta$,  H$\delta$}   (4.6)     &  10.0&3.62 $\pm$ 0.01&0.27$\pm$ 0.01     \\ 
  UDF-2383 &        53.14642 &       -27.77831 &      1.865  $\pm$
  0.001  &      24.701  $\pm$       0.009        &     0.239  $\pm$
  0.067&    ETG&   O~III     &&&      \\
 UDF-2433 &        53.14744 &       -27.77761 &        1.825  $\pm$
 0.002&      22.708  $\pm$       0.003       & 1.459  $\pm$
 0.056 &          ETG&  O~III,  {\it O~IIIx}     &  10.4&2.5 $\pm$ 0.1&6.8$\pm$ 0.1        \\
   UDF-2491 &        53.14744 &       -27.77761 &       1.889   $    \pm$       0.007 &             & 1.459  $\pm$       0.056 &     LTG&   O~III     &&&  \\    
       UDF-2798 &        53.13924 &       -27.77485 &      1.843  $\pm$       0.002  &      25.689  $\pm$       0.051         & 0.442  $\pm$       0.115  &   ETG&  OII, O~III    &&&      \\
            UDF-2900 &        53.15288 &       -27.77250 &       1.845  $\pm$       0.001 &      24.019  $\pm$       0.005
            & 0.481  $\pm$       0.043   &ETG&O~III, H$\beta$, {\it
              H$\gamma$   }  &  9.8&0.74 $\pm$ 0.01&1.33$\pm$ 0.03          \\
 UDF-3035$^{a}$ &        53.15605 &       -27.77095 &        1.833  $\pm$       0.004 &      24.587  $\pm$       0.009    & 0.889  $\pm$       0.085 &    LTG&    {\it OII, O~III, H$\gamma$}      &&&   \\
            UDF-3045 &        53.15228 &       -27.77009 &      1.848  $\pm$       0.001  &      23.144  $\pm$       0.003 &0.743  $\pm$       0.029     &LTG&O~III, {\it H$\beta$, H$\gamma$}    &  10.0&2.18 $\pm$ 0.01&1.53$\pm$ 0.04       \\
           UDF-3058 &        53.16001 &       -27.77100 &       1.853
           $\pm$       0.002 &      25.223  $\pm$       0.012     &
           0.381  $\pm$       0.072  &    ETG&    O~III   &&&                  \\
            UDF-3510 &        53.16387 &       -27.76532 &        1.841  $\pm$       0.003 &      24.241  $\pm$       0.007      & 0.494  $\pm$       0.046 &   LTG&     O~III, {\it H$\beta$, H$\gamma$}    &  9.4&4.10 $\pm$ 0.05&0.79$\pm$ 0.01       \\
GMASS220&53.15492&-27.80940&1.850 $\pm$       0.001 &&&LTG&Fe, C (4.7) &&&\\ 
\tableline \tableline\\
HUDFJ0332.5-2747.3&&&&&\\
\tableline \\
       UDF-463 &        53.15881 &       -27.79716 &       1.904
         $\pm$       0.004 &      24.292  $\pm$       0.008      &
         2.239  $\pm$       0.062   & ETG&4000\AA break &  10.8&0.66 $\pm$ 0.01&2.49$\pm$ 0.02  \\
             UDF-669 &        53.14060 &       -27.79562 &       1.909
             $\pm$       0.002 &      23.970  $\pm$       0.005      &
             0.730  $\pm$       0.029   & ETG&O~III, H$\beta$, {\it
               H$\delta$}    &  10.8&0.66 $\pm$ 0.01&2.49$\pm$ 0.02    \\
            UDF-1180 &        53.14930 &       -27.78853 &       1.907
            $\pm$       0.002 &      22.464  $\pm$       0.002      &
            0.862  $\pm$       0.020   & LTG&O~III, OII, H$\beta$,  {\it O~IIIx,
   H$\delta$}    &  10.1&2.56 $\pm$ 0.02&1.42$\pm$ 0.01       \\
            UDF-1355 &        53.14799 &       -27.78769 &       1.884
            $\pm$       0.006 &      22.475  $\pm$       0.002      &
            0.805  $\pm$       0.038   & ETG&O~III  (4.4)     &  9.7&1.90 $\pm$ 0.01&0.69$\pm$ 0.01    \\
            UDF-1698 &        53.16935 &       -27.78499 &       1.911
            $\pm$       0.002 &      24.701  $\pm$       0.009      &
            -0.150  $\pm$       0.196   & ETG&O~III   &&&\\
         {\it UDF-1898$^{a}$}  &        53.17368 &       -27.78207 &       1.894
          $\pm$       0.058 &      22.708  $\pm$       0.003      &
          2.140  $\pm$       0.228   & LTG&{\it O~IIIx, H$\beta$}   &&&\\
           {\it UDF-1909$^{a}$}  &        53.14903 &       -27.78196 &       1.917
            $\pm$       0.017 &      24.489  $\pm$       0.008      &
            1.050  $\pm$       0.041   &LTG& {\it O~III, H$\delta$ } &&&  \\
            UDF-2797 &        53.14418 &       -27.77356 &       1.892
            $\pm$       0.001 &      23.624  $\pm$       0.005      &
            0.385  $\pm$       0.027   &LTG& O~III,  H$\beta$,  H$\delta$,
            {\it  H$\gamma$}  &  9.8&0.74 $\pm$ 0.01&0.37$\pm$ 0.01  \\
            UDF-3297 &        53.14102 &       -27.76673 &       1.904
            $\pm$       0.006 &      23.189  $\pm$       0.004      &
            2.246  $\pm$       0.036   & ETG&4000\AA break \\
\tableline \tableline\\
\end{tabular}}
\end{center}
\small{
{\bf Note.} Galaxies are identified by their 3D-HST ID (UDF; from Brammer et al. 2012) or GMASS ID (GMASS; from Kurk et al. 2013). 
Redshifts and uncertainties for GMASS are from Kurk et al. (2013). Magnitudes and colors are from Guo et al. (2013). The two galaxy pairs
are not separate in this catalog, and we give the magnitude of the Guo
et al. object for one of the two galaxies.
For the spectral features,
O~III and O~IIIx 
indicate the  [O~III]$\lambda$5007 and the  [O~III]$\lambda$4363 emission lines, respectively, H$\beta$ the [H$\beta$]$\lambda$4861
emission line,   H$\gamma$ the [H$\gamma$]$\lambda$4340, and OII the [O~II]$\lambda3727$. The lines
are in italics if they were measured with a $1<S/N < 3$, otherwise they were measured with $S/N > 3$. When the galaxies also have a GMASS redshift, in parenthesis
is given the GMASS redshift $S/N$. The flag (a) means all lines
were measured with $S/N<1$. 
Galaxies with
IDs in italic are not considered as structure members because they have low $S/N$ spectroscopy, and are not detected in the $336W$ bandpass. We estimated structural properties for all galaxies with $H_{160} < 24.5$~mag, where the $J_{125}$ images were available. Statistical uncertainties on masses are a few dex, while systematics, due to the use of different spectral energy distribution templates are $<0.5$~dex (e.g. Delaye et al. 2014). The uncertainties on galaxy sizes and the Sersic index are the fit uncertainties given by GALFIT. The typical systematic uncertainties on $Re$, $\sqrt{b/a}$and Sersic index at these magnitudes are $\sim 20\%$ up to $H_{160} = 24.5$~mag (e.g. van der Wel et al. 2012; Basset et al. 2013).
}
\end{table*}

\section{DISCUSSION AND CONCLUSIONS}

Deep mid--infrared surveys, and space and ground--based
infrared spectroscopy have enabled the discovery of clusters
of galaxies at redshift $z=1.5-2$, an epoch largely unexplored until recently. 
Most of these discoveries have been based on the searches for star--forming
galaxy overdensities around radio sources, and/or red galaxy
overdensities in the mid--infrared with Spitzer IRAC. The advent of the HST WFC3 grism and ground--based infrared spectroscopy
permits confirmation of these discoveries as real galaxy
overdensities (Stanford et al. 2012; Zeimann et al. 2012; Gobat et
al. 2013; Newman et al. 2013). 

Current X--ray and SZ observations probe cluster
virialization through the detection of the hot gas in the gravitational potential well, down
to cluster masses of $\approx 10^{14} M_\odot$ and up to redshift
$z \approx 1$. At higher redshifts, only the extreme end of the cluster mass function can be detected by current instruments.
A few objects at $1.5<z<2$ correspond to significant X--ray 
detections and were identified as already virialized
(Andreon et al. 2009; Gobat et al. 2011; Santos et
al. 2011; Stanford et al. 2012; Mantz et al. 2014). Two of them also show a significant SZ signal (Brodwin et al. 2012; Mantz et al. 2014). Their cluster masses cover the range of $M_{200}\approx (0.5 - 4)  \times 10^{14}
M_{\sun}$. The other detections (e.g., less massive objects) can only currently be identified as
significant red galaxy overdensities, without confirmation of 
virialization by the detection of hot gas. Depending on the presence,
or not, of the red sequence and their richness, these objects have been
identified as clusters or proto--clusters (e.g., Pentericci et
al. 2000; Miley et al. 2004, 2006; Venemans 2007; Kuiper et al. 2010; Hatch et al. 2011).

In this paper, we presented the discovery of two star--forming galaxy overdensities in the HUDF using HST WFC3 grism spectroscopy and imaging
observations from the CANDELS and 3D-HST Treasury programs. The richest overdensity,  HUDFJ0332.4-2746.6, includes 18 spectroscopic members, of which 6 are
ETGs. The other one, HUDFJ0332.5-2747.3,
includes 7 spectroscopic members, of which 3 are ETGs. Our
detections are mostly based on line emitter galaxy overdensities, similar to current proto--cluster discoveries at $z>2$, but different from current cluster detections at the same redshift that are based on red galaxy overdensities.
We confirmed the grism redshifts using deep  far-UV photometry from the UVUDF (Teplitz et al. 2013).

\begin{table*}
\begin{center}
\caption{Comparison of HUDFJ0332.4-2746.6  and HUDFJ0332.5-2747.3 properties with those of already known clusters, proto--clusters and groups at $z=1.6-2$ \label{tab2}}
\vspace{0.2cm}
\resizebox{!}{2cm}{
\begin{tabular}{llccccccccccccccc}
\tableline \tableline\\
Name &Identification &z&Overdensity&$\sigma_{disp}$&Mass&X--ray Lum./Detection&Reference\\
&& &&(km/s)&($10^{14} \times M_\odot $)&($10^{43}$ erg s$^{-1}$)&\\
\tableline \tableline \\
CL J033211.67-274633.8 &Group&1.61&$\sim 5\sigma$&...&$M_{200}^{(a)} =0.32\pm0.08$&$1.8\pm0.6$&Tanaka et al. \\
IRC-0218A/XMM-LSS J02182-05102&Proto--cluster&1.62&$>20\sigma$&$860\pm490$&$M_{200}^{(b)} \sim 0.1-0.4$&$>4\sigma$ Detection&Papovich et al. 2010; 2012\\
SpARCS J022427-032354&Cluster&1.63&...&...&...&Detection&Muzzin et al. (2013)\\
IDCS J1426+3508&Cluster&1.75&...&...&$M_{200}^{(a)}\sim5.6\pm1.6$&$55\pm12$ &Stanford et al. 2012; Brodwin et al. 2012\\
JKCS~041&Cluster&1.80&...&...&$M_{200}^{(c)}\sim2$&$76\pm5$& Newman et al. 2013; Andreon et al. 2013\\
HUDFJ0332.4-2746.6 &Proto--cluster&1.84&$\sim 20\sigma$&$730\pm 260$&$M_{200}^{(b)} = 2.2 \pm 1.8 $&$<1-6$& This work\\
IDCS J1433.2+3306&Cluster&1.89&...&...&$M_{200}\sim 1$&...&Zeimann et al. 2012\\
HUDFJ0332.5-2747.3 &Group&1.90&$\sim 4-7\sigma$&...&...&...& This work\\
CL J1449+085&Cluster&1.99&$>20\sigma$&...&$M_{200}^{(a)} =0.53\pm0.09$&$6.4\pm1.8$&Gobat et al. 2013\\
\tableline \tableline\\
\end{tabular}}
\end{center}
\small{{\bf Note.} All estimates are given as they are from the references. For the overdensities, $\sigma$ is estimated with respect to the background, as given by the references. X--ray fluxes and mass estimates have not been homogenized. (a) and (b) indicate mass estimates derived from the X--ray flux and the velocity dispersion, respectively. (c) indicates that the mass estimate is derived from the X--ray flux and cluster richness.}
\end{table*}

 Using a Nth-nearest neighbor distance estimator and the density contrast, we measure a galaxy overdensity at $\sim 20 \sigma$ and $\sim (4-7) \sigma$ above the background, for HUDFJ0332.4-2746.6 and HUDFJ0332.5-2747.3, respectively. Under the hypothesis of viralization, from HUDFJ0332.4-2746.6 velocity dispersion, we obtain a mass estimate of $M_{200}=(2.2  \pm 1.8)  \times 10^{14}
M_{\sun}$, consistent with the lack of
extended X--ray emission.  In Table~3, we compare our newly discovered structure to already known clusters, proto--clusters and groups at $z=1.6-2$. Within the uncertainties, HUDFJ0332.4-2746.6 has the properties characteristic of a proto--cluster, because of its overdensity and estimated mass, and  HUDFJ0332.5-2747.3 those of a galaxy group, because of its overdensity. 

Predictions from numerical simulations (Cohn et White 2005; Li et
al. 2007; Chiang et al. 2013; Cautun et al. 2014) suggest that HUDFJ0332.4-2746.6 is most probably a progenitor of $M_{200}\approx 10^{14}
M_{\sun}$ galaxy clusters at $z\sim 1$ and of $M_{200}\approx few  \times 10^{14}
M_{\sun}$ galaxy clusters at the present. At $z\approx 1.8-1.9$ Chiang et
al. (2013) predict the comoving effective sizes of clusters of mass $M_{200}\approx 10^{14}
M_{\sun}$  to be $\approx 2-5$~Mpc. Their total mass extends beyond this spatial scale, based on the 
cosmological N--body simulation from the Millennium Run (Springel et
al. 2005) and semi-analytic galaxy catalogs from Guo et
al. (2011). 

Within the GOODS--CDFS area covered by the Guo et al. (2013) photometric redshift catalog, we searched for overdensities in photometric redshift ranges around the two overdensities and found several groups. Without extensive spectroscopic follow--up we cannot conclude that these groups are at the same spectroscopic redshift as our newly discovered structures. It would be interesting to follow them up spectroscopically and understand if our two overdensities are part of a larger structure at the same redshift.

We estimate that at most $\approx 50\%$ of the proto--cluster
members are ETGs, against the $80\%$ observed in
clusters of galaxies at $z \approx 1-1.5$ (e.g., Postman et al. 2005; Mei et
al. 2009; Mei et al. 2012).  About $50\%$ of the
structure members show possible interactions or disturbed morphologies
(asymmetries, faint substructures, and tails), which are possible signatures of merger remnants or disk instability.  This suggests mergers and possibly disk instabilities as the primary and ongoing mechanisms of assembly in at least half of the galaxies in dense environments at these redshifts.

  For galaxy clusters and proto--clusters at $z=1.6-1.9$, the ETG fractions can be quite different in different objects, going from $50\%$ (Gobat 2013; Zeimann et al. 2012; Muzzin et al. 2013) to $80\%$ (Papovich et al. 2012). The lower end of these estimated fractions and our results are close to the fractions of ETGs with mass of $M>10^{10}  M_{\sun}$ obtained from Mortlock et al. (2013) in the CANDELS Ultra--Deep Survey (UDS). This suggests the existence of significant overdensities that have similar ETG fractions as the field. It is also interesting that Mortlock et al. found that $z\sim 1.85$ is a redshift of transition between an epoch in which irregular galaxy fractions dominate over disk galaxy fractions to an epoch in which the trend is inverted to the type fractions observed in the local Universe.

Using multi--wavelength photometry from Guo et al. (2013),
we study the two structures' galaxy colors, and find that their red sequence is not yet in place. All the
confirmed ETG members, but two, show emission lines that indicate
recent star formation activity. Only one ETG shows colors
consistent with  those characteristic of an old stellar population at
these redshifts, e.g., all the others have active stellar
populations. This is consistent with the fact that most of the ETGs in
the two structures are star--forming and will be quenched only at a later
time. 

 From both of the two structures' ETG fractions and their colors, new ETGs would need to be formed (e.g., by transformations of LTGs by environmental effects; e.g., Boselli \& Gavazzi 2006) or 
accreted, to obtain the higher ETG fractions observed at lower redshifts. The progenitors of some of these newly transformed ETGs could have been observed as a  passive bulge--dominated LTG population in clusters and dense regions at $z=1-1.3$ (Bundy et al. 2010; Mei et al. 2006ab, 2012; George et al. 2013).

Current red sequence galaxies are predicted to form the bulk of
their stars at an average formation redshift of $z_f=2-3$ from both the interpretation of their scaling relations  and
age and metallicity measurements (e.g., Thomas et al. 2005), and
semianalytic models based on the Millennium simulation (e.g., De Lucia
\& Blaizot 2007; Barro et al. 2013b; Shankar et al. 2013).
This implies that part of their progenitors at $z\approx2$ are star--forming
galaxies. Combined deep
high resolution space imaging and grism spectroscopy permitted us to spectroscopically confirm
 star--forming blue ETG progenitors. At least part of the red
sequence ETGs are already ETGs and are compact before quenching their star
formation. Our results are consistent with recent observations in the HUDF and modeling by
Barro et al. (2013a,b) that demonstrated how 
compact star--forming galaxies (all morphology selected) appear to be progressively quenched from $z=2-3$ to
$z=1-2$. In this work, we spectroscopically confirm for the first time the presence of star--forming blue compact ETGs
in significant galaxy overdensities, e.g. in a proto--cluster. Since star--forming ETGs are rare both in clusters and the field up to $z\approx 1.5$ (e.g., Mei et al. 2009; Huertas--Company et al. 2010; Brodwin et al. 2013; Barro et
al. 2013ab, and references therein), the star--forming ETGs
are most probably (at least part of) the progenitors of passive ETGs in galaxy clusters at
$z\sim 1-1.5$.

We compare the masses and the sizes of  the structures' star--forming blue ETGs with those of passive ETGs in dense regions and galaxy clusters at $z=1-2$, and find that they lie on the same mass--size relation.  Interestingly, quiescent ETGs in galaxy clusters at $z=1.8=2$ show a similar behavior as our structurer's blue star--forming ETGs, and the mass-normalized B--band rest-frame size, $\gamma$, does not significantly evolve in the redshift range $0.7<z<2$, contrary to field ETGs (Damjanov et al. 2011; Cimatti et al. 2012; Newman et al. 2013).
This implies that, if these objects are the progenitors of quiescent ETGs in clusters at $z=1-1.5$, their mass--size relation did not evolve significantly even if their star--formation was quenched; galaxies could increase their mass, simultaneously increasing their size according to this relation. 

The diversity of these structures shows how overdensities at $z>1.5$ have less homogeneous galaxy populations than those at $z<1.5$. Large studies of clusters and proto--clusters at these higher redshift have to quantify how detection techniques impact their sample selection function, to obtain good statistics of their galaxy population.

\section{SUMMARY}

We found star--forming blue ETGs in two newly discovered galaxy overdensities at $z=1.84$ and $z=1.9$ in the HUDF. We summarize our main results here:

\begin{itemize}

\item We discovered two galaxy overdensities in the HUDF. The first is identified as a galaxy proto--cluster at $z = 1.84 \pm 0.01$, HUDFJ0332.4-2746.6, and includes 18 spectroscopic members, for a galaxy overdensity of $\sim 20\sigma$. The second is a galaxy group at $z = 1.90 \pm 0.01$, HUDFJ0332.5-2747.3, with seven spectroscopic members, and a galaxy overdensity of $\sim  4-7\sigma$. Under the hypothesis of viralization, from its velocity dispersion, we obtain a mass estimate for  HUDFJ0332.4-2746.6 of $M_{200}=(2.2  \pm 1.8) \times 10^{14} M_{\sun}$, consistent with the lack of extended X--ray emission.

\item The two structures have not yet formed a red sequence. For the first time, we confirm a significant presence of star--forming blue ETGs in dense environments at $z\sim1.8-1.9$. We classified eight and five ETGs in HUDFJ0332.4-2746.6 and HUDFJ0332.5-2747.3, respectively, of which five have $J_{125}<24.5$~mag. The ETG fraction in both structures is at most $\sim 50\%$, similar to fractions obtained in some galaxy clusters (Gobat 2013; Zeimann et al. 2012; Muzzin et al. 2013) and close to those obtained in the field at these redshifts (Mortlock et al. 2013). These are lower fractions than what is observed in some other galaxy clusters at similar redshifts (Papovich et al. 2012) and in galaxy clusters at $z<1.5$ ($80\%$; e.g., Postman et al. 2005; Desai et al. 2007; Mei et
al. 2009; Mei et al. 2012). This suggests that large overdensities at $z>1.5$ have more diverse galaxy populations than those at $z<1.5$, and that it is essential to quantify how detection techniques impact our cluster/proto--cluster selection function. 

\item About $50\%$ of the structure members show possible interactions or disturbed morphologies, with asymmetries, faint substructures, and tails, all possible signatures of merger remnants or disk instabilities.  This suggests mergers and possibly disk instabilities as the primary and ongoing mechanisms of assembly in at least half of the galaxies in dense environments at these redshifts.

\item The star--forming blue ETG have masses of $8.9 \lesssim log_{10} (\frac{M}{M_{\sun}}) \lesssim 10.8$, and their mass--size relation lies on the same mass--size relation observed for quiescent ETGs in clusters and dense regions at $z=0.7-2$ (Lani et al. 2013; Newman et al. 2013; Strazzullo et al. 2013; Delaye et al. 2014). Interestingly,  quiescent ETG sizes in clusters also do not evolve significantly in this redshift range, which covers $\sim 4$~Gyr in time. This suggest that at these epochs, cluster ETGs do not  significantly change their median/average sizes, and evolve according to a mass--size relation similar to the one at $z\sim1$.

\item Both of the two structures' ETG fractions and their colors suggest that these star--forming blue ETGs are the most likely progenitors of at least part of the passive ETGs observed in clusters at $z<1$. Their masses are $\sim 3-5 $ times lower than the most massive ETGs in these lower redshift clusters. More (massive) ETGs have to be formed/accreted and then quenched, to obtain the ETG fractions, colors, and masses observed in clusters at $z<1$. 

\end{itemize}

Small samples can hardly be representative of the larger populations, but
as with other studies at these high redshifts, we discover new objects
often one by one, and we are consistently building
larger samples that will improve our understanding of
cluster formation and evolution. The
CANDELS and 3D-HST Treasury programs have opened a new path for
proto--cluster detection in this redshift range.

Surveys of this kind point to the capabilities of future space
missions,  such as Euclid (Laureijs et al. 2011) and WFIRST (Thompson et al. 2013). 
Those missions have the potential to discover a large
population of young clusters at all redshifts, and
especially at these very early epochs of cluster formation and assembly.



\acknowledgments This work is based on observations taken by the CANDELS Multi-Cycle Treasury Program and the 3D-HST Treasury Program (GO 12177 and 12328) with the NASA/ESA HST, which is operated by the Association of Universities for Research in Astronomy, Inc., under NASA contract NAS5-26555. This work is based in part on observations made with the Spitzer Space Telescope, which is operated by the Jet Propulsion Laboratory, California Institute of Technology under a contract with NASA. S.M. acknowledges financial support from the Institut Universitaire de France (IUF), of which she is senior member. We thank the referee for her/his very useful comments that improved the paper.



{\it Facilities:} \facility{HST(ACS and WFC3)},  \facility{Spitzer (IRAC)}

\clearpage



\clearpage








\appendix{
\section{Appendix}

In this appendix, we show detections and dropouts from the UVUDF survey
(Tepliz et al. 2013). Most of the selected galaxies are $F225W$ dropouts, e.g., are not detected in  WFC3 $F225W$ (the top left panel), but are detected in WFC3 $F336W$ (the top right panel). The two exceptions are:  UDF~1909 and UDF~1898, which are not detected in $F336W$, but are detected in ACS $F435W$, and are most probably at higher redshift galaxies or have too low surface brightness to be unambiguously identified as $z=1.8-1.9$ galaxies.

\begin{figure}
\epsscale{.80}
\plotone{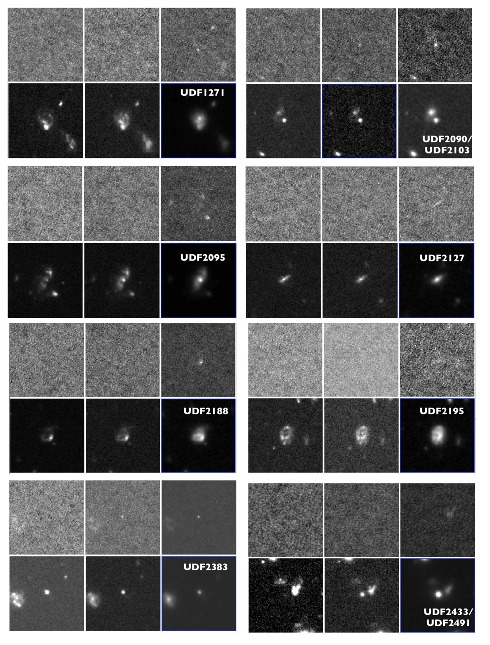}
\caption{HUDFJ0332.4-2746.6 candidates. For each candidate, starting from the top left,
we show clockwise WFC3 $F225W$, $F275W$ and
$F336W$ from UVUDF, ACS $F435W$, $I_{814}$ and WFC3 $J_{125}$ images.  Galaxies are identified by their 3D-HST ID. The size of each image is 5\arcsec. All candidates are $F275W$ dropouts. \label{uvudf11}}
\end{figure}

\begin{figure}
\epsscale{.80}
\plotone{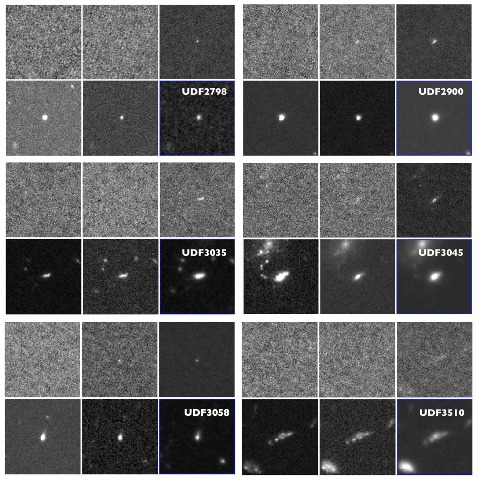}
\caption{ HUDFJ0332.4-2746.6 candidates. For each candidate, starting from the top left,
we show clockwise WFC3 $F225W$, $F275W$ and
$F336W$ from UVUDF, ACS $F435W$, $I_{814}$ and WFC3 $J_{125}$ images.  Galaxies are identified by their 3D-HST ID. The size of each image is 5\arcsec.  All candidates are $F275W$ dropouts. \label{uvudf12}}
\end{figure}

\begin{figure}
\epsscale{.80}
\plotone{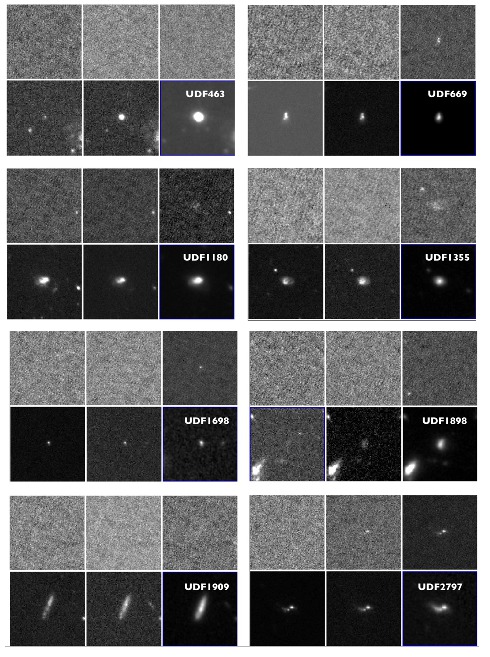}
\caption{HUDFJ0332.5-2747.3 candidates\label{uvudf22}. For each candidate, starting from the top left,
we show clockwise WFC3 $F225W$, $F275W$ and
$F336W$ from UVUDF, ACS $F435W$, $I_{814}$ and WFC3 $J_{125}$ images.  Galaxies are identified by their 3D-HST ID. The size of each image is 5\arcsec.  All candidates are $F275W$ dropouts, except for UDF~1909 and UDF~1898. Those two galaxies are not detected in F336W, but are detected in ACS F435W, e.g. they are F435W dropouts. They are excluded from our analysis because they are most probably higher redshift galaxies or have too low surface brightness to be unambiguously identified as $z=1.8-1.9$ galaxies.}
\end{figure}
}





\begin{thebibliography}{}



\bibitem[Anderson et al.(2015)]{2015MNRAS.449.3806A} Anderson, M.~E., 
Gaspari, M., White, S.~D.~M., Wang, W., \& Dai, X.\ 2015, \mnras, 449, 3806 

\bibitem[Andreon et 
al.(2009)]{2009A&A...507..147A} Andreon, S., Maughan, B., Trinchieri, G., \& Kurk, J.\ 2009, \aap, 507, 147 


\bibitem[Andreon 
\& Huertas-Company(2011)]{2011A&A...526A..11A} Andreon, S., \& Huertas-Company, M.\ 2011, \aap, 526, A11 


\bibitem[Andreon et 
al.(2014)]{2014A&A...565A.120A} Andreon, S., Newman, A.~B., Trinchieri, G., et al.\ 2014, \aap, 565, A120 




\bibitem[Arnouts et al.(2002)]{2002MNRAS.329..355A} Arnouts, S., 
Moscardini, L., Vanzella, E., et al.\ 2002, \mnras, 329, 355 

\bibitem[Ashby et al.(2009)]{2009ApJ...701..428A} Ashby, M.~L.~N., Stern, 
D., Brodwin, M., et al.\ 2009, \apj, 701, 428 


\bibitem[Ashby et al.(2013)]{2013ApJS..209...22A} Ashby, M.~L.~N., 
Stanford, S.~A., Brodwin, M., et al.\ 2013a, \apjs, 209, 22 


\bibitem[Ashby et al.(2013)]{2013ApJ...769...80A SEDS} Ashby, M.~L.~N., Willner, 
S.~P., Fazio, G.~G., et al.\ 2013b, \apj, 769, 80 




\bibitem[Barro et al.(2013)]{2013ApJ...765..104B} Barro, G., Faber, S.~M., 
P{\'e}rez-Gonz{\'a}lez, P.~G., et al.\ 2013, \apj, 765, 104 



\bibitem[Barro et al.(2014)]{2014ApJ...791...52B} Barro, G., Faber, S.~M., 
P{\'e}rez-Gonz{\'a}lez, P.~G., et al.\ 2014, \apj, 791, 52 


\bibitem[Bassett et al.(2013)]{2013ApJ...770...58B} Bassett, R., Papovich, 
C., Lotz, J.~M., et al.\ 2013, \apj, 770, 58 




\bibitem[Beckwith et al.(2006)]{2006AJ....132.1729B} Beckwith, S.~V.~W., 
Stiavelli, M., Koekemoer, A.~M., et al.\ 2006, \aj, 132, 1729 


\bibitem[Bernardi et al.(2014)]{2014MNRAS.443..874B} Bernardi, M., Meert, 
A., Vikram, V., et al.\ 2014, \mnras, 443, 874 




\bibitem[Bertin 
\& Arnouts(1996)]{1996A&AS..117..393B} Bertin, E., \& Arnouts, S.\ 1996, \aaps, 117, 393 


\bibitem[Bird 
\& Beers(1993)]{1993AJ....105.1596B} Bird, C.~M., \& Beers, T.~C.\ 1993, \aj, 105, 1596 


\bibitem[Bouwens et al.(2011)]{2011ApJ...737...90B} Bouwens, R.~J., 
Illingworth, G.~D., Oesch, P.~A., et al.\ 2011, \apj, 737, 90 

\bibitem[Boselli \& Gavazzi(2006)]{bo06} 
Boselli, A., \& Gavazzi, G.\ 2006, \pasp, 118, 517 



\bibitem[Brammer et al.(2012)]{2012ApJS..200...13B} Brammer, G.~B., van 
Dokkum, P.~G., Franx, M., et al.\ 2012, \apjs, 200, 13 


\bibitem[Brodwin et al.(2012)]{2012ApJ...753..162B} Brodwin, M., Gonzalez, 
A.~H., Stanford, S.~A., et al.\ 2012, \apj, 753, 162 


\bibitem[Brodwin et al.(2013)]{2013ApJ...779..138B} Brodwin, M., Stanford, 
S.~A., Gonzalez, A.~H., et al.\ 2013, \apj, 779, 138 





\bibitem[Bruzual A.~\& Charlot(2003)]{BC2003}
Bruzual A., G.~\& Charlot, S.\ 2003, \mnras, 344, 1000 (BC03)


\bibitem[Bundy et al.(2010)]{2010ApJ...719.1969B} Bundy, K., Scarlata, C., 
Carollo, C.~M., et al.\ 2010, \apj, 719, 1969 



\bibitem[Calzetti et al.(2000)]{ca00} Calzetti, D., Armus, 
L., Bohlin, R.~C., Kinney, A.~L., Koornneef, J., 
\& Storchi-Bergmann, T.\ 2000, \apj, 533, 682 


\bibitem[Carlberg et al.(1997)]{1997ApJ...485L..13C} Carlberg, R.~G., Yee, 
H.~K.~C., Ellingson, E., et al.\ 1997, \apjl, 485, L13 



\bibitem[Castellano et al.(2007)]{2007ApJ...671.1497C} Castellano, M., 
Salimbeni, S., Trevese, D., et al.\ 2007, \apj, 671, 1497 




\bibitem[Castellano et 
al.(2011)]{2011A&A...530A..27C} Castellano, M., Pentericci, L., Menci, N., et al.\ 2011, \aap, 530, A27 


\bibitem[Cautun et al.(2014)]{2014MNRAS.441.2923C} Cautun, M., van de 
Weygaert, R., Jones, B.~J.~T., \& Frenk, C.~S.\ 2014, \mnras, 441, 2923 



\bibitem[Chabrier(2003)]{2003PASP..115..763C} Chabrier, G.\ 2003, \pasp, 
115, 763 


\bibitem[Chiaberge et al.(2010)]{2010ApJ...710L.107C} Chiaberge, M., 
Capetti, A., Macchetto, F.~D., et al.\ 2010, \apjl, 710, L107 




\bibitem[Chiang et al.(2013)]{2013ApJ...779..127C} Chiang, Y.-K., Overzier, 
R., \& Gebhardt, K.\ 2013, \apj, 779, 127 



\bibitem[Cimatti et al.(2012)]{2012MNRAS.422L..62C} Cimatti, A., Nipoti, 
C., \& Cassata, P.\ 2012, \mnras, 422, L62 




\bibitem[Colbert et al.(2013)]{2013ApJ...779...34C} Colbert, J.~W., 
Teplitz, H., Atek, H., et al.\ 2013, \apj, 779, 34 



\bibitem[Cohn 
\& White(2005)]{2005APh....24..316C} Cohn, J.~D., \& White, M.\ 2005, Astroparticle Physics, 24, 316 



\bibitem[Conselice(2003)]{2003ApJS..147....1C} Conselice, C.~J.\ 2003, 
\apjs, 147, 1 




\bibitem[Cooper et al.(2008)]{2008MNRAS.383.1058C} Cooper, M.~C., Newman, 
J.~A., Weiner, B.~J., et al.\ 2008, \mnras, 383, 1058 


\bibitem[Cooper et al.(2012)]{2012MNRAS.419.3018C} Cooper, M.~C., Griffith, 
R.~L., Newman, J.~A., et al.\ 2012, \mnras, 419, 3018 



\bibitem[Damjanov et al.(2011)]{2011ApJ...739L..44D} Damjanov, I., Abraham, 
R.~G., Glazebrook, K., et al.\ 2011, \apjl, 739, L44 



\bibitem[Danese et 
al.(1980)]{1980A&A....82..322D} Danese, L., de Zotti, G., \& di Tullio, G.\ 1980, \aap, 82, 322 


\bibitem[De Lucia 
\& Blaizot(2007)]{2007MNRAS.375....2D} De Lucia, G., \& Blaizot, J.\ 2007, \mnras, 375, 2 



\bibitem[Delaye et al.(2014)]{2014MNRAS.441..203D} Delaye, L., 
Huertas-Company, M., Mei, S., et al.\ 2014, \mnras, 441, 203 


\bibitem[Desai et al.(2007)]{2007ApJ...660.1151D} Desai, V., Dalcanton, 
J.~J., Arag{\'o}n-Salamanca, A., et al.\ 2007, \apj, 660, 1151 




\bibitem[Dressler(1980)]{1980ApJ...236..351D} Dressler, A.\ 1980, \apj, 
236, 351 




\bibitem[Eisenhardt et al.(2008)]{2008ApJ...684..905E} Eisenhardt, 
P.~R.~M., Brodwin, M., Gonzalez, A.~H., et al.\ 2008, \apj, 684, 905 



\bibitem[Elbaz et 
al.(2007)]{2007A&A...468...33E} Elbaz, D., Daddi, E., Le Borgne, D., et al.\ 2007, \aap, 468, 33 


\bibitem[Ellis et al.(2013)]{2013ApJ...763L...7E} Ellis, R.~S., McLure, 
R.~J., Dunlop, J.~S., et al.\ 2013, \apjl, 763, L7 



\bibitem[Evrard et al.(2008)]{2008ApJ...672..122E} Evrard, A.~E., Bialek, 
J., Busha, M., et al.\ 2008, \apj, 672, 122 




\bibitem[Fassbender et 
al.(2011)]{2011A&A...527L..10F} Fassbender, R., Nastasi, A., B{\"o}hringer, H., et al.\ 2011, \aap, 527, L10 



\bibitem[Fazio et al.(2004)]{2004ApJS..154...10F} Fazio, G.~G., Hora, 
J.~L., Allen, L.~E., et al.\ 2004, \apjs, 154, 10 

\bibitem[Fern{\'a}ndez Lorenzo et al.(2013)]{2013MNRAS.434..325F} 
Fern{\'a}ndez Lorenzo, M., Sulentic, J., Verdes-Montenegro, L., 
\& Argudo-Fern{\'a}ndez, M.\ 2013, \mnras, 434, 325 




\bibitem[Galametz et al.(2012)]{2012ApJ...749..169G} Galametz, A., Stern, 
D., De Breuck, C., et al.\ 2012, \apj, 749, 169 




\bibitem[George et al.(2013)]{2013ApJ...770..113G} George, M.~R., Ma, 
C.-P., Bundy, K., et al.\ 2013, \apj, 770, 113 




\bibitem[Gladders 
\& Yee(2000)]{2000AJ....120.2148G} Gladders, M.~D., \& Yee, H.~K.~C.\ 2000, \aj, 120, 2148 



\bibitem[Gobat et 
al.(2011)]{2011A&A...526A.133G} Gobat, R., Daddi, E., Onodera, M., et al.\ 2011, \aap, 526, A133 


\bibitem[Gobat et al.(2013)]{2013ApJ...776....9G} Gobat, R., Strazzullo, 
V., Daddi, E., et al.\ 2013, \apj, 776, 9 



\bibitem[Grogin et al.(2011)]{2011ApJS..197...35G} Grogin, N.~A., Kocevski, 
D.~D., Faber, S.~M., et al.\ 2011, \apjs, 197, 35 


\bibitem[Gr{\"u}tzbauch et al.(2011)]{2011MNRAS.418..938G} Gr{\"u}tzbauch, 
R., Conselice, C.~J., Bauer, A.~E., et al.\ 2011, \mnras, 418, 938 



\bibitem[Guo et al.(2013)]{2013ApJS..207...24G} Guo, Y., Ferguson, H.~C., 
Giavalisco, M., et al.\ 2013, \apjs, 207, 24 



\bibitem[Hahn et al.(2007)]{2007MNRAS.375..489H} Hahn, O., Porciani, C., 
Carollo, C.~M., \& Dekel, A.\ 2007a, \mnras, 375, 489 

\bibitem[Hahn et al.(2007)]{2007MNRAS.381...41H} Hahn, O., Carollo, C.~M., 
Porciani, C., \& Dekel, A.\ 2007b, \mnras, 381, 41 



\bibitem[Hatch et al.(2011)]{2011MNRAS.415.2993H} Hatch, N.~A., Kurk, 
J.~D., Pentericci, L., et al.\ 2011, \mnras, 415, 2993 




\bibitem[Hayashi et al.(2011)]{2011MNRAS.415.2670H} Hayashi, M., Kodama, 
T., Koyama, Y., Tadaki, K.-I., \& Tanaka, I.\ 2011, \mnras, 415, 2670 

\bibitem[Hatch et al.(2011)]{2011MNRAS.410.1537H} Hatch, N.~A., De Breuck, 
C., Galametz, A., et al.\ 2011, \mnras, 410, 1537 



\bibitem[Huertas-Company et 
al.(2009)]{2009A&A...497..743H} Huertas-Company, M., Tasca, L., Rouan, D., et al.\ 2009, \aap, 497, 743 

\bibitem[Huertas-Company et 
al.(2010)]{2010A&A...515A...3H} Huertas-Company, M., Aguerri, J.~A.~L., Tresse, L., et al.\ 2010, \aap, 515, A3 




\bibitem[Huertas-Company et 
al.(2011)]{2011A&A...525A.157H} Huertas-Company, M., Aguerri, J.~A.~L., Bernardi, M., Mei, S., \& S{\'a}nchez Almeida, J.\ 2011, \aap, 525, A157 


\bibitem[Huertas-Company et al.(2013)]{2013MNRAS.428.1715H} 
Huertas-Company, M., Mei, S., Shankar, F., et al.\ 2013a, \mnras, 428, 1715 



\bibitem[Huertas-Company et al.(2013)]{2013ApJ...779...29H} 
Huertas-Company, M., Shankar, F., Mei, S., et al.\ 2013b, \apj, 779, 29 



\bibitem[Ichikawa et al.(2006)]{2006SPIE.6269E..16I} Ichikawa, T., Suzuki, 
R., Tokoku, C., et al.\ 2006, \procspie, 6269, 626916 




\bibitem[Ilbert et 
al.(2006)]{2006A&A...457..841I} Ilbert, O., Arnouts, S., McCracken, H.~J., et al.\ 2006, \aap, 457, 841 

\bibitem[Ilbert et al.(2010)]{2010ApJ...709..644I} Ilbert, O., Salvato, M., 
Le Floc'h, E., et al.\ 2010, \apj, 709, 644 

\bibitem[Illingworth et al.(2013)]{2013ApJS..209....6I} Illingworth, G.~D., 
Magee, D., Oesch, P.~A., et al.\ 2013, \apjs, 209, 6 




\bibitem[Kartaltepe et al.(2014)]{2014arXiv1401.2455K} Kartaltepe, J.~S., 
Mozena, M., Kocevski, D., et al.\ 2014, arXiv:1401.2455 




\bibitem[Kodama et al.(2007)]{2007MNRAS.377.1717K} Kodama, T., Tanaka, I., 
Kajisawa, M., et al.\ 2007, \mnras, 377, 1717 



\bibitem[Koekemoer et al.(2011)]{2011ApJS..197...36K} Koekemoer, A.~M., 
Faber, S.~M., Ferguson, H.~C., et al.\ 2011, \apjs, 197, 36 


\bibitem[Koekemoer et al.(2013)]{2013ApJS..209....3K} Koekemoer, A.~M., 
Ellis, R.~S., McLure, R.~J., et al.\ 2013, \apjs, 209, 3 




\bibitem[Koyama et al.(2013)]{2013MNRAS.tmp.1677K} Koyama, Y., Smail, I., 
Kurk, J., et al.\ 2013, \mnras, 1677 



\bibitem[Kuiper et al.(2010)]{2010MNRAS.405..969K} Kuiper, E., Hatch, 
N.~A., R{\"o}ttgering, H.~J.~A., et al.\ 2010, \mnras, 405, 969 



\bibitem[K{\"u}mmel et al.(2009)]{2009PASP..121...59K} K{\"u}mmel, M., 
Walsh, J.~R., Pirzkal, N., Kuntschner, H., 
\& Pasquali, A.\ 2009, \pasp, 121, 59 




\bibitem[Kuntschner et al.(2010)]{2010SPIE.7731E.104K} Kuntschner, H., 
Bushouse, H., K{\"u}mmel, M., Walsh, J.~R., 
\& MacKenty, J.\ 2010, \procspie, 7731,  

\bibitem[Kurk et 
al.(2009)]{2009A&A...504..331K} Kurk, J., Cimatti, A., Zamorani, G., et al.\ 2009, \aap, 504, 331 


\bibitem[Kurk et 
al.(2013)]{2013A&A...549A..63K} Kurk, J., Cimatti, A., Daddi, E., et al.\ 2013, \aap, 549, A63 


\bibitem[Laidler et al.(2007)]{2007PASP..119.1325L} Laidler, V.~G., 
Papovich, C., Grogin, N.~A., et al.\ 2007, \pasp, 119, 1325 


\bibitem[Lani et al.(2013)]{2013MNRAS.435..207L} Lani, C., Almaini, O., 
Hartley, W.~G., et al.\ 2013, \mnras, 435, 207 


\bibitem[Laureijs et al.(2011)]{2011arXiv1110.3193L} Laureijs, R., Amiaux, 
J., Arduini, S., et al.\ 2011, arXiv:1110.3193 



\bibitem[Li et al.(2007)]{2007MNRAS.379..689L} Li, Y., Mo, H.~J., van den 
Bosch, F.~C., \& Lin, W.~P.\ 2007, \mnras, 379, 689 



\bibitem[Liu et al.(2013)]{2013ApJ...769..147L} Liu, F.~S., Guo, Y., Koo, 
D.~C., et al.\ 2013, \apj, 769, 147 


\bibitem[Lotz et al.(2013)]{2013ApJ...773..154L} Lotz, J.~M., Papovich, C., 
Faber, S.~M., et al.\ 2013, \apj, 773, 154 



\bibitem[MacKenty(2012)]{2012SPIE.8442E..1VM} MacKenty, J.~W.\ 2012, 
\procspie, 8442, 84421V 



\bibitem[Mantz et al.(2014)]{2014ApJ...794..157M} Mantz, A.~B., Abdulla, 
Z., Carlstrom, J.~E., et al.\ 2014, \apj, 794, 157 


\bibitem[McLean et al.(2008)]{2008SPIE.7014E..2ZM} McLean, I.~S., Steidel, 
C.~C., Matthews, K., Epps, H., 
\& Adkins, S.~M.\ 2008, \procspie, 7014, 70142Z 


\bibitem[McLean et al.(2012)]{2012SPIE.8446E..0JM} McLean, I.~S., Steidel, 
C.~C., Epps, H.~W., et al.\ 2012, \procspie, 8446, 84460J 




\bibitem[Mei et al.(2006)]{2006ApJ...639...81M} Mei, S., Blakeslee, J.~P., 
Stanford, S.~A., et al.\ 2006a, \apj, 639, 81 




\bibitem[Mei et al.(2006)]{2006ApJ...644..759M} Mei, S., Holden, B.~P., 
Blakeslee, J.~P., et al.\ 2006b, \apj, 644, 759 



\bibitem[Mei et al.(2009)]{2009ApJ...690...42M} Mei, S., Holden, B.~P., 
Blakeslee, J.~P., et al.\ 2009, \apj, 690, 42 


\bibitem[Mei et al.(2012)]{2012ApJ...754..141M} Mei, S., Stanford, S.~A., 
Holden, B.~P., et al.\ 2012, \apj, 754, 141 


\bibitem[Merrall 
\& Henriksen(2003)]{2003ApJ...595...43M} Merrall, T.~E.~C., \& Henriksen, R.~N.\ 2003, \apj, 595, 43 



\bibitem[Miley et al.(2004)]{2004Natur.427...47M} Miley, G.~K., Overzier, 
R.~A., Tsvetanov, Z.~I., et al.\ 2004, \nat, 427, 47 


\bibitem[Miley et al.(2006)]{2006ApJ...650L..29M} Miley, G.~K., Overzier, 
R.~A., Zirm, A.~W., et al.\ 2006, \apjl, 650, L29 





\bibitem[Mortlock et al.(2013)]{2013MNRAS.433.1185M} Mortlock, A., 
Conselice, C.~J., Hartley, W.~G., et al.\ 2013, \mnras, 433, 1185 



\bibitem[Munari et al.(2013)]{2013MNRAS.430.2638M} Munari, E., Biviano, A., 
Borgani, S., Murante, G., \& Fabjan, D.\ 2013, \mnras, 430, 2638 




\bibitem[Muzzin et al.(2013)]{2013ApJ...767...39M} Muzzin, A., Wilson, G., 
Demarco, R., et al.\ 2013, \apj, 767, 39 

\bibitem[Nakamura(2000)]{2000ApJ...531..739N} Nakamura, T.~K.\ 2000, \apj, 
531, 739 




\bibitem[Navarro et al.(1996)]{1996ApJ...462..563N} Navarro, J.~F., Frenk, 
C.~S., \& White, S.~D.~M.\ 1996, \apj, 462, 563 


\bibitem[Newman et al.(2012)]{2012ApJ...746..162N} Newman, A.~B., Ellis, 
R.~S., Bundy, K., \& Treu, T.\ 2012, \apj, 746, 162 




\bibitem[Newman et al.(2013)]{2013arXiv1310.6754N} Newman, A.~B., Ellis, 
R.~S., Andreon, S., et al.\ 2013, arXiv:1310.6754 



\bibitem[Oke 
\& Gunn(1983)]{1983ApJ...266..713O} Oke, J.~B., \& Gunn, J.~E.\ 1983, \apj, 266, 713 




\bibitem[Papovich et al.(2010)]{2010ApJ...716.1503P} Papovich, C., 
Momcheva, I., Willmer, C.~N.~A., et al.\ 2010, \apj, 716, 1503 



\bibitem[Papovich et al.(2012)]{2012ApJ...750...93P} Papovich, C., Bassett, 
R., Lotz, J.~M., et al.\ 2012, \apj, 750, 93 


\bibitem[Peng, Ho, Impey, \& Rix(2002)]{pe02} 
Peng, C.~Y., Ho, L.~C., Impey, C.~D., \& Rix, H.\ 2002, \aj, 124, 266 


\bibitem[Peng et al.(2010)]{2010AJ....139.2097P} Peng, C.~Y., Ho, L.~C., 
Impey, C.~D., \& Rix, H.-W.\ 2010, \aj, 139, 2097 



\bibitem[Pentericci et 
al.(2000)]{2000A&A...361L..25P} Pentericci, L., Kurk, J.~D., R{\"o}ttgering, H.~J.~A., et al.\ 2000, \aap, 361, L25 



\bibitem[Poggianti et al.(2013)]{2013ApJ...762...77P} Poggianti, B.~M., 
Calvi, R., Bindoni, D., et al.\ 2013, \apj, 762, 77 




\bibitem[Popesso et 
al.(2012)]{2012A&A...537A..58P} Popesso, P., Biviano, A., Rodighiero, G., et al.\ 2012, \aap, 537, A58 


\bibitem[Postman et al. (2005)]{po05}
Postman, M. et al., 2005, ApJ, 623, 721



\bibitem[Raichoor et al.(2012)]{2012ApJ...745..130R} Raichoor, A., Mei, S., 
Stanford, S.~A., et al.\ 2012, \apj, 745, 130 



\bibitem[Rettura et al.(2010)]{2010ApJ...709..512R} Rettura, A., Rosati, 
P., Nonino, M., et al.\ 2010, \apj, 709, 512 



\bibitem[Rigby et al.(2013)]{2013MNRAS.tmp.2671R} Rigby, E.~E., Hatch, 
N.~A., R{\"o}ttgering, H.~J.~A., et al.\ 2013, \mnras, 2671 


\bibitem[Rykoff et al.(2008)]{2008MNRAS.387L..28R} Rykoff, E.~S., Evrard, 
A.~E., McKay, T.~A., et al.\ 2008, \mnras, 387, L28 



\bibitem[Salimbeni et 
al.(2009)]{2009A&A...501..865S} Salimbeni, S., Castellano, M., Pentericci, L., et al.\ 2009, \aap, 501, 865 


\bibitem[Santos et 
al.(2011)]{2011A&A...531L..15S} Santos, J.~S., Fassbender, R., Nastasi, A., et al.\ 2011, \aap, 531, L15 


\bibitem[Santos et al.(2014)]{2014MNRAS.438.2565S} Santos, J.~S., Altieri, 
B., Tanaka, M., et al.\ 2014, \mnras, 438, 2565 



\bibitem[Scoville et al.(2013)]{2013ApJS..206....3S} Scoville, N., Arnouts, 
S., Aussel, H., et al.\ 2013, \apjs, 206, 3 

\bibitem[Shankar et al.(2013)]{2013MNRAS.428..109S} Shankar, F., Marulli, 
F., Bernardi, M., et al.\ 2013, \mnras, 428, 109 


\bibitem[Shankar et al.(2014)]{2014MNRAS.tmp..344S} Shankar, F., Mei, S., 
Huertas-Company, M., et al.\ 2014, \mnras, 344 



\bibitem[Sharples et al.(2006)]{2006NewAR..50..370S} Sharples, R., Bender, 
R., Bennett, R., et al.\ 2006, \nar, 50, 370 


\bibitem[Shattow et al.(2013)]{2013MNRAS.433.3314S} Shattow, G.~M., Croton, 
D.~J., Skibba, R.~A., et al.\ 2013, \mnras, 433, 3314 




\bibitem[Sirianni et al.(2005)]{2005PASP..117.1049S} Sirianni, M., Jee, 
M.~J., Ben{\'{\i}}tez, N., et al.\ 2005, \pasp, 117, 1049 



\bibitem[Snyder et al.(2012)]{2012ApJ...756..114S} Snyder, G.~F., Brodwin, 
M., Mancone, C.~M., et al.\ 2012, \apj, 756, 114 




\bibitem[Stanford et al.(2012)]{2012ApJ...753..164S} Stanford, S.~A., 
Brodwin, M., Gonzalez, A.~H., et al.\ 2012, \apj, 753, 164 


\bibitem[Steidel et al.(1998)]{1998ApJ...492..428S} Steidel, C.~C., 
Adelberger, K.~L., Dickinson, M., et al.\ 1998, \apj, 492, 428 




\bibitem[Strazzullo et al.(2013)]{2013ApJ...772..118S} Strazzullo, V., 
Gobat, R., Daddi, E., et al.\ 2013, \apj, 772, 118 


\bibitem[Tadaki et al.(2012)]{2012MNRAS.423.2617T} Tadaki, K., Kodama, 
T., Ota, K., et al.\ 2012, \mnras, 423, 2617 


\bibitem[Tanaka et al.(2010)]{2010ApJ...716L.152T} Tanaka, M., Finoguenov, 
A., \& Ueda, Y.\ 2010, \apjl, 716, L152 


\bibitem[Tanaka et al.(2013)]{2013PASJ...65...17T} Tanaka, M., Finoguenov, 
A., Mirkazemi, M., et al.\ 2013, \pasj, 65, 17 



\bibitem[Teplitz et al.(2013)]{2013AJ....146..159T} Teplitz, H.~I., 
Rafelski, M., Kurczynski, P., et al.\ 2013, \aj, 146, 159 

\bibitem[Thomas et al.(2005)]{2005ApJ...621..673T} Thomas, D., Maraston, 
C., Bender, R., \& Mendes de Oliveira, C.\ 2005, \apj, 621, 673 



\bibitem[Thompson et al.(2013)]{2013arXiv1312.4548T} Thompson, R., Green, 
J., Rieke, G., et al.\ 2013, arXiv:1312.4548 



\bibitem[Tran et al.(2010)]{2010ApJ...719L.126T} Tran, K.-V.~H., Papovich, 
C., Saintonge, A., et al.\ 2010, \apjl, 719, L126 




\bibitem[van Dokkum et al.(2013)]{2013arXiv1305.2140V} van Dokkum, P., 
Brammer, G., Momcheva, I., et al.\ 2013, arXiv:1305.2140 


\bibitem[van der Wel et al.(2012)]{2012ApJS..203...24V} van der Wel, A., 
Bell, E.~F., H{\"a}ussler, B., et al.\ 2012, \apjs, 203, 24 


\bibitem[Venemans et 
al.(2007)]{2007A&A...461..823V} Venemans, B.~P., R{\"o}ttgering, H.~J.~A., Miley, G.~K., et al.\ 2007, \aap, 461, 823 


\bibitem[Voit(2005)]{2005RvMP...77..207V} Voit, G.~M.\ 2005, Reviews of 
Modern Physics, 77, 207 



\bibitem[Vulcani et 
al.(2013)]{2013A&A...550A..58V} Vulcani, B., Poggianti, B.~M., Oemler, A., et al.\ 2013, \aap, 550, A58 



\bibitem[White et al.(2010)]{2010MNRAS.408.1818W} White, M., Cohn, J.~D., 
\& Smit, R.\ 2010, \mnras, 408, 1818 


\bibitem[Wuyts et al.(2008)]{2008ApJ...682..985W} Wuyts, S., Labb{\'e}, I., 
Schreiber, N.~M.~F., et al.\ 2008, \apj, 682, 985 


\bibitem[Wuyts et al.(2009)]{2009ApJ...706..885W} Wuyts, S., van Dokkum, 
P.~G., Franx, M., et al.\ 2009, \apj, 706, 885 


\bibitem[Wylezalek et al.(2013)]{2013ApJ...769...79W} Wylezalek, D., 
Galametz, A., Stern, D., et al.\ 2013, \apj, 769, 79 


\bibitem[Yang et al.(2007)]{2007ApJ...671..153Y} Yang, X., Mo, H.~J., van 
den Bosch, F.~C., et al.\ 2007, \apj, 671, 153 



\bibitem[Yuan et al.(2014)]{2014ApJ...795L..20Y} Yuan, T., Nanayakkara, T., 
Kacprzak, G.~G., et al.\ 2014, \apjl, 795, LL20 




\bibitem[Xue et al.(2011)]{2011ApJS..195...10X} Xue, Y.~Q., Luo, B., 
Brandt, W.~N., et al.\ 2011, \apjs, 195, 10 




\bibitem[Zeimann et al.(2012)]{2012ApJ...756..115Z} Zeimann, G.~R., 
Stanford, S.~A., Brodwin, M., et al.\ 2012, \apj, 756, 115 

\bibitem[Ziparo et al.(2014)]{2014MNRAS.437..458Z} Ziparo, F., Popesso, P., 
Finoguenov, A., et al.\ 2014, \mnras, 437, 458 





\end{thebibliography}
\end{document}